\documentclass{aa}

\usepackage[varg]{txfonts}
\usepackage{siunitx}
\usepackage[version=4]{mhchem}
\usepackage{multirow}
\usepackage{graphicx}
\usepackage[caption=false]{subfig}
\usepackage[strict]{changepage}
\usepackage{xcolor}
\usepackage{ulem}
\usepackage[colorlinks=true,citecolor=blue]{hyperref}
\usepackage{float}
\usepackage{amssymb}
\usepackage{amsmath}
\usepackage{lscape}
\usepackage{mathtools}
\usepackage{natbib}
\usepackage{pdflscape}
\usepackage{gensymb}

\defcitealias{zhou2023}{Paper~I}   

\DeclareSIUnit\jansky{Jy}

\newcolumntype{L}[1]{>{\raggedright\let\newline\\\arraybackslash\hspace{0pt}}m{#1}}
\newcolumntype{C}[1]{>{\centering\let\newline\\\arraybackslash\hspace{0pt}}m{#1}}
\newcolumntype{R}[1]{>{\raggedleft\let\newline\\\arraybackslash\hspace{0pt}}m{#1}}

\begin{document}

\title{The star formation histories, star formation efficiencies and ionizing sources of ATLASGAL clumps with HII regions}
\author{J. W. Zhou\inst{\ref{inst1}} 
\and Sami Dib \inst{\ref{inst2}}
\and Pavel Kroupa\inst{\ref{inst3}}
\fnmsep \inst{\ref{inst4}}
}
\institute{
Max-Planck-Institut f\"{u}r Radioastronomie, Auf dem H\"{u}gel 69, 53121 Bonn, Germany \label{inst1} \\
\email{jwzhou@mpifr-bonn.mpg.de}
\and
Max Planck Institute f\"{u}r Astronomie, K\"{o}nigstuhl 17, 69117 Heidelberg, Germany \label{inst2}\\
\email{dib@mpia.de}
\and
Helmholtz-Institut f{\"u}r Strahlen- und Kernphysik (HISKP), Universität Bonn, Nussallee 14–16, 53115 Bonn, Germany \label{inst3}\\
\email{pkroupa@uni-bonn.de}
\and
Charles University in Prague, Faculty of Mathematics and Physics, Astronomical Institute, V Hole{\v s}ovi{\v c}k{\'a}ch 2, CZ-180 00 Praha 8, Czech Republic \label{inst4}
}

\date{Accepted XXX. Received YYY; in original form ZZZ}

\abstract
{1226 ATLASGAL clumps with HII regions (HII-clumps) were matched with radio sources in the CORNISH-North/South surveys, and 392 of them have corresponding radio sources. We determined the stellar luminosity $L_\mathrm{*, T84}$ according to the Lyman continuum flux $N_\mathrm{Ly}$.  When the bolometric luminosity of HII-clumps is less than log$_{10}$(L$_{\rm bol, obs}$/L$_{\odot}$) $\approx$ 3.7, corresponding to a clump mass log$_{10}$(M$_{\rm cl}$/M$_{\odot}$) $\approx$ 2.55, the values of $L_\mathrm{*, T84}$ derived from $N_\mathrm{Ly}$ overestimate the actual stellar luminosities, because the accretion onto the protostars contributes significantly to the radio emission. After subtracting the  accretion luminosity from $L_\mathrm{*, T84}$, we obtained reasonable estimates of the stellar luminosity. Using the 0.5 Myr isochrone, we calculated the stellar masses according to the stellar luminosities, and found that they roughly follow the $m_{\rm max}-M_{\rm ecl}$ relation of embedded clusters, consistent with the ionizing sources representing the most massive stars in the embedded clusters of HII-clumps. We also studied the contribution of the possible flaring activity to the observed stellar luminosity and found that they can be neglected. We further studied the change of SFE with the clump mass. According to the derived mass of the most massive star in each HII-clump, using the theoretical $m_{\rm max}-M_{\rm ecl}$ relation, we calculated the mass of the corresponding embedded cluster and then the SFE of the clump. The SFE decreases with  increasing clump mass, with a median value of $\approx$0.3. We also independently derived the SFE for each HII-clump based on the model developed in our previous work. The SFEs of HII-clumps derived from the observation and the model are in good agreement. 
Concerning the star formation histories of the ATLASGAL clumps, low-mass clumps may reach the peak of star formation earlier than high-mass clumps, consistent with the shorter free-fall time of low-mass clumps.
}


\titlerunning{Star formation properties of ATLASGAL clumps}
\authorrunning{J. W. Zhou, Sami Dib, Pavel Kroupa}

\maketitle 

\section{Introduction}
Understanding the contribution of stars in different mass ranges to the total energetics and dynamics of star clusters is crucial.
A majority, if not all, of the stars we observe were born within embedded clusters \citep{Kroupa1995a-277, Kroupa1995b-277, Lada2003-41,Kroupa2005-576,Megeath2016-151, Dinnbier2022-660}, which adds significant intricacy to pinpointing and analyzing individual protostellar objects. 
Unlike their lower-mass counterparts, massive stars evolve towards the main sequence while still deeply embedded within their parental clumps. This renders their initial stages invisible, even when observed using mid-infrared wavelengths. Nowadays,
our comprehension of massive star and cluster formation has significantly advanced, owing to the combination of Galactic plane surveys and high-angular resolution images obtained through submillimeter facilities \citep{Motte2018-56}. Surveys such as HiGAL \citep{Molinari2010} and ATLASGAL \citep{Schuller2009-504} have played pivotal roles by providing unbiased compilations of dense clumps that track the early phases of massive star and cluster formation. Explorations employing representative clump samples can provide valuable insights into the efficiency of converting molecular gas into stellar clusters \citep{Elia2021-504,Urquhart2022-510,Wells2022-516}.

In \citet{Urquhart2022-510}, 5007 ATLASGAL clumps have been categorized into four distinct evolutionary stages, with the most advanced stage being referred to as HII-clumps, i.e. ATLASGAL clumps with HII regions. 
In \citet{Zhou2024arXiv240809867Z} (paper I), we synthesized the embedded clusters within HII-clumps under the assumption that the stellar initial mass function (IMF) follows a universal optimal distribution function rather than a probability density \citep{Kroupa2013-115,Yan2023-670}.  
We utilized the 0.1 Myr isochrone to estimate the bolometric luminosity of individual stars within an embedded cluster, augmenting this with the accretion luminosity of each star in the cluster. The cumulative bolometric luminosity of the synthetic embedded clusters aligns closely with the observed bolometric luminosity of HII-clumps, validating the effectiveness of the method. As a follow up study based on paper I, in this work, we focus on the ionizing sources in HII-clumps. They may represent the most massive stars in the embedded clusters. We explore their physical properties by comparing the observations with numerical simulations.

\section{Sample}
\begin{figure}
\centering
\includegraphics[width=0.47\textwidth]{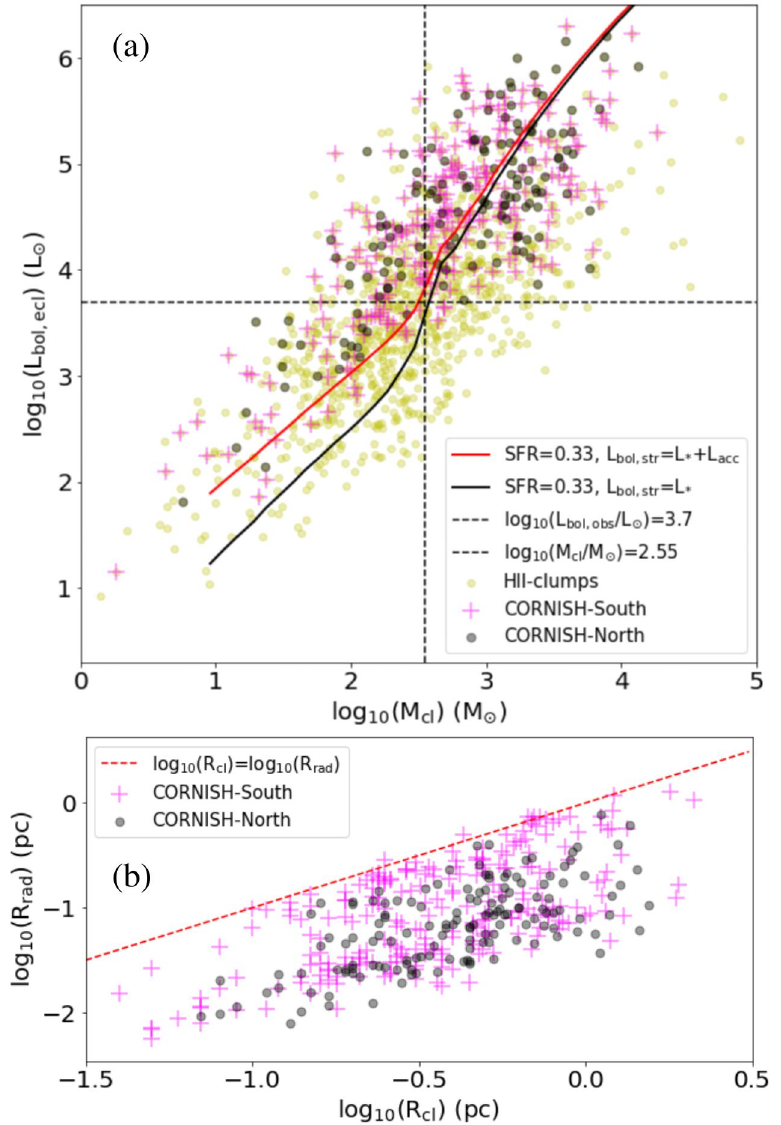}
\caption{(a) The L$_{\rm bol, obs}$-M$_{\rm cl}$ relation of HII-clumps. Black dots and magenta pluses mark the HII-clumps matched with radio sources in the CORNISH-North and CORNISH-South surveys. The red and black lines show the fitting to the observed HII-clumps using the optimal sampling model in \citet{Zhou2024arXiv240809867Z}. The red line considers the accretion luminosity of each star in an embedded cluster, the black line does not. The vertical and horizontal lines mark the turning points of the bolometric luminosity and mass of HII-clumps discussed in Sec.\ref{turning};
(b) The radii of radio sources (R$_{\rm rad}$) vs. the radii of the corresponding HII-clumps (R$_{\rm cl}$).}
\label{sample}
\end{figure}

The physical parameters of 1246 HII-clumps have been calculated and listed in \citet{Urquhart2022-510}. 1226 HII-clumps with mass ($M_{\rm cl}$) and bolometric luminosity ($L_{\rm bol, obs}$) measurements were matched with radio sources in the catalogs of the 
CORNISH-North \citep{Purcell2013-205}
and the CORNISH-South \citep{Irabor2023-520} surveys. The separations between the central coordinates of a radio source and a HII-clump was required to be smaller than the radius of the clump. 
Generally, one HII-clump matches with one radio source. If one HII-clump includes more than one radio source, we only consider the radio source with the strongest radio emission, because we are only interested in the most massive star in a HII-clump. Finally, the numbers of radio sources matched with HII-clumps are 244 and 148 in the CORNISH-South and the CORNISH-North survey catalogs, respectively.

Fig.~\ref{sample}(a) displays the samples of HII clumps in the $L_{\rm bol, obs}$-$M_{\rm cl}$ diagram. We can see that the samples are mainly concentrated at the upper end, similar to Fig.~5 in \citet{Urquhart2022-510}, where the HII-clumps were divided into two populations, i.e. radio-loud and radio-quiet. Our samples are radio-loud, which are more luminous than the radio-quiet HII-clumps. This indicates that they are hosting high-mass stars.
In Fig.~\ref{sample}(b), the radii of radio sources are systematically smaller than the radii of the HII-clumps. These radio sources should be excited by the most massive stars in the embedded clusters of HII-clumps. From the radio emission, we can infer some physical parameters of the massive stars.

\section{Results and Discussion}

\begin{table*}
\centering
\caption{Explanation of the physical quantities.}
\label{tab1}
\begin{tabular}{cc}
\hline

$L_{\rm bol, obs}$ & The bolometric luminosity of a HII-clump \\
$M_{\rm cl}$ & The mass of a HII-clump \\
$M_{\rm ecl}$ & The mass in stars of an embedded cluster in a HII-clump \\
$N_\mathrm{Ly}$ & The Lyman continuum flux \\
$L_\mathrm{*, T84}$ & The luminosity of a star calculated by interpolating the star's luminosity given as \\ 
& a function of the Lyman continuum flux in Table. 1 of \citet{Thompson1984-283} \\
$m_\mathrm{*, iso}$ & The mass of a star estimated by the 0.5 Myr isochrone according to $L_\mathrm{*, T84}$ \\
$m_{\rm max}$ & The mass of the most massive star in a synthetic embedded cluster generated by the GalIMF code \\
$L_\mathrm{acc}$ & The accretion luminosity of the most massive star in a synthetic embedded cluster\\
$L_\mathrm{*}$ & The cleaned luminosity of a star after subtracting its accretion luminosity from $L_\mathrm{*, T84}$ \\
$L_{\rm x}$ and $L_{\rm opt}$ & The luminosities of a flare in X-ray and optical bands \\

\hline
\label{para}
\end{tabular}
\end{table*}

\subsection{Lyman continuum flux}\label{Lyman}
\begin{figure}
\centering
\includegraphics[width=0.47\textwidth]{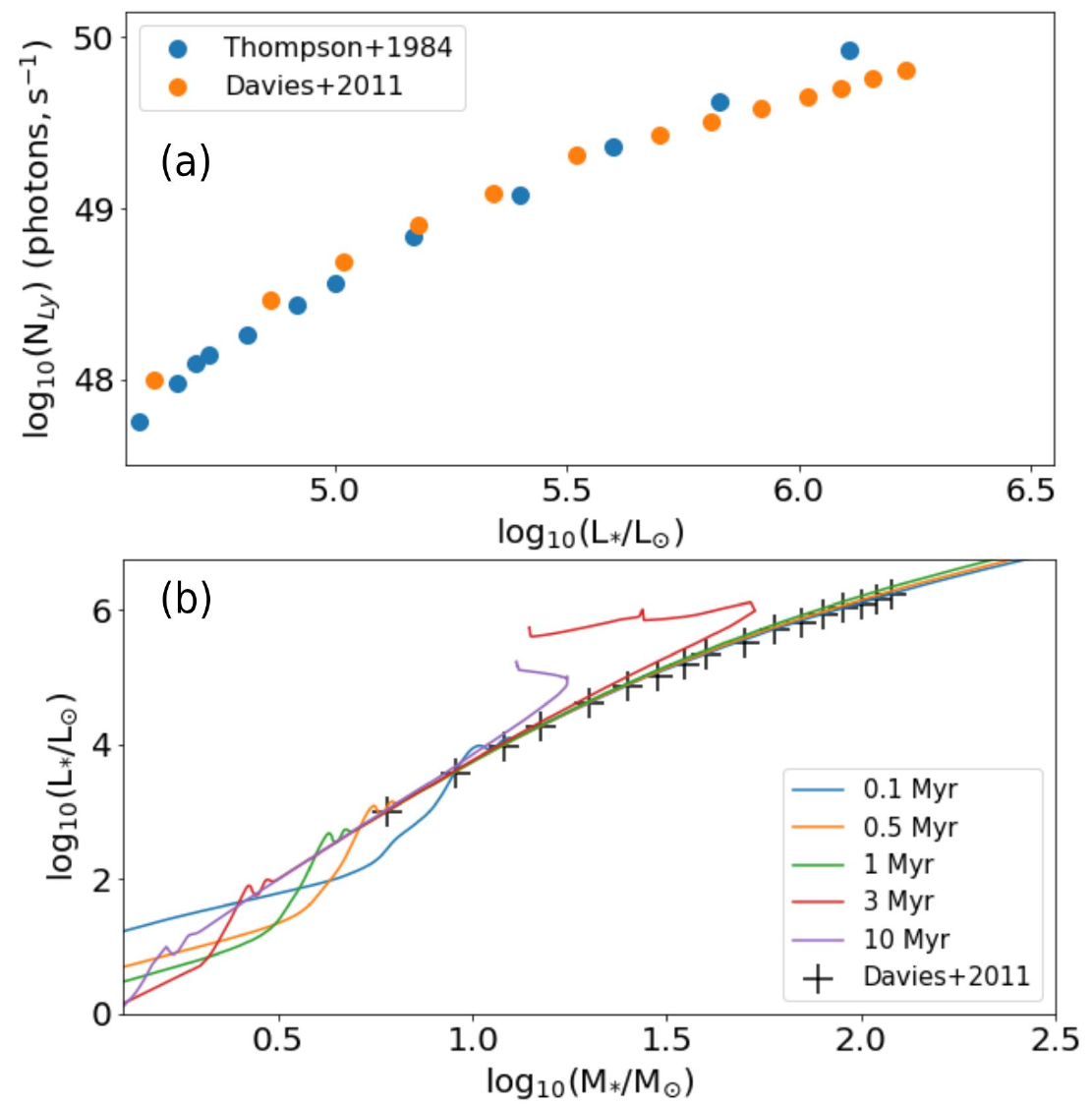}
\caption{(a) The relationships between stellar luminosity $L_\mathrm{*}$ of individual stars and Lyman continuum flux $N_\mathrm{Ly}$ in Table. 1 of \citet{Davies2011-416} (orange dots) and Table. 1 of \citet{Thompson1984-283} (blue dots), respectively; (b) The relationship between
stellar mass and luminosity in Table. 1 of \citet{Davies2011-416}. The color-coded lines are the MESA isochrones, see text for details.}
\label{iso}
\end{figure}

As discussed in \citet{Kalcheva2018-615}, assuming optically thin radio sources at 5 GHz (CORNISH-North) and 5.5 GHz (CORNISH-South) is justified for our samples. We calculated the physical parameters of the matched radio sources following the formalism of \citet{Mezger1967-147} and \citet{Rubin1968-154} using the equations collected in \citet{Sanchez-Monge2013-550}.
The emission measure ($EM$) and the electron density ($n_\mathrm{e}$) were calculated using:
\begin{equation}
\bigg[\frac{EM}{\mathrm{cm}^{-6}~\mathrm{pc}}\bigg]\,=\,
1.7\times10^7~\bigg[\frac{S_\nu}{\mathrm{Jy}}\bigg]
\bigg[\frac{\nu}{\mathrm{GHz}}\bigg]^{0.1}
\bigg[\frac{T_\mathrm{e}}{\mathrm{K}}\bigg]^{0.35}
\bigg[\frac{\theta_\mathrm{s}}{\arcsec}\bigg]^{-2},
\label{em}
\end{equation}
and
\begin{equation}
\bigg[\frac{n_\mathrm{e}}{\mathrm{cm}^{-3}}\bigg]\,=\,
2.3\times10^6~\bigg[\frac{S_\nu}{\mathrm{Jy}}\bigg]^{0.5}
\bigg[\frac{\nu}{\mathrm{GHz}}\bigg]^{0.05}
\bigg[\frac{T_\mathrm{e}}{\mathrm{K}}\bigg]^{0.175}
\bigg[\frac{d}{\mathrm{pc}}\bigg]^{-0.5}
\bigg[\frac{\theta_\mathrm{s}}{\arcsec}\bigg]^{-1.5},
\label{ne}
\end{equation}
where $S_\nu$ is the integrated radio flux density, $T_\mathrm{e}$ is the electron temperature assumed to be $10^4$~K, $\theta_\mathrm{s}$ is the angular diameter of the source, $d$ is the distance, and $\nu$ is the observing frequency.
$S_\nu$ and $\theta_\mathrm{s}$ are assigned the values in the catalogs of \citet{Purcell2013-205} and \citet{Irabor2023-520}. 
The values of $d$ are the distances of the corresponding HII-clumps listed in \citet{Urquhart2022-510}.
The number of Lyman-continuum photons per second ($N_\mathrm{Ly}$; hereafter Lyman continuum flux) is calculated from the flux density and distance as:
%
\begin{equation}
\bigg[\frac{N_\mathrm{Ly}}{\mathrm{photons} 
\, \mathrm{s}^{-1}}\bigg]\,=\,
8.9\times10^{40}~\bigg[\frac{S_\nu}{\mathrm{Jy}}\bigg]
\bigg[\frac{\nu}{\mathrm{GHz}}\bigg]^{0.1}
\bigg[\frac{T_\mathrm{e}}{10^4\mathrm{K}}\bigg]^{-0.45}
\bigg[\frac{d}{\mathrm{pc}}\bigg]^2.
\label{nl}
\end{equation}

Table.~1 of \citet{Davies2011-416} gives the relationship between the stellar parameters and Lyman continuum flux.
For stars with masses above 60 M$_{\odot}$, they made the assumption that Lyman continuum flux per unit area is constant, which is likely to slightly underestimate the Lyman flux of very massive stars.
In Fig.~\ref{iso}(a), we compared the data in Table. 1 of \citet{Davies2011-416} with Table. 1 of \citet{Thompson1984-283}. 
It seems that Table.~1 of \citet{Thompson1984-283} preforms better at the end with high $N_\mathrm{Ly}$ and stellar luminosity, $L_\mathrm{*}$. The relationship between
stellar mass and luminosity in \citet{Davies2011-416} was taken from \citet{Meynet2000-361}, which is consistent with the isochrones of 0.5 Myr or 1 Myr from the MESA Isochrones and Stellar Tracks (MIST) project \citep{Choi2016-823}, as shown in Fig.~\ref{iso}(b).

In this work, we determined the stellar luminosity $L_\mathrm{*, T84}$ corresponding to the calculated $N_\mathrm{Ly}$ by interpolating the
stellar luminosity given as a function of the Lyman continuum flux in Table.~1 of \citet{Thompson1984-283}, shown in Fig.~\ref{iso}(a).

\subsection{The turning points}\label{turning}
\begin{figure}
\centering
\includegraphics[width=0.47\textwidth]{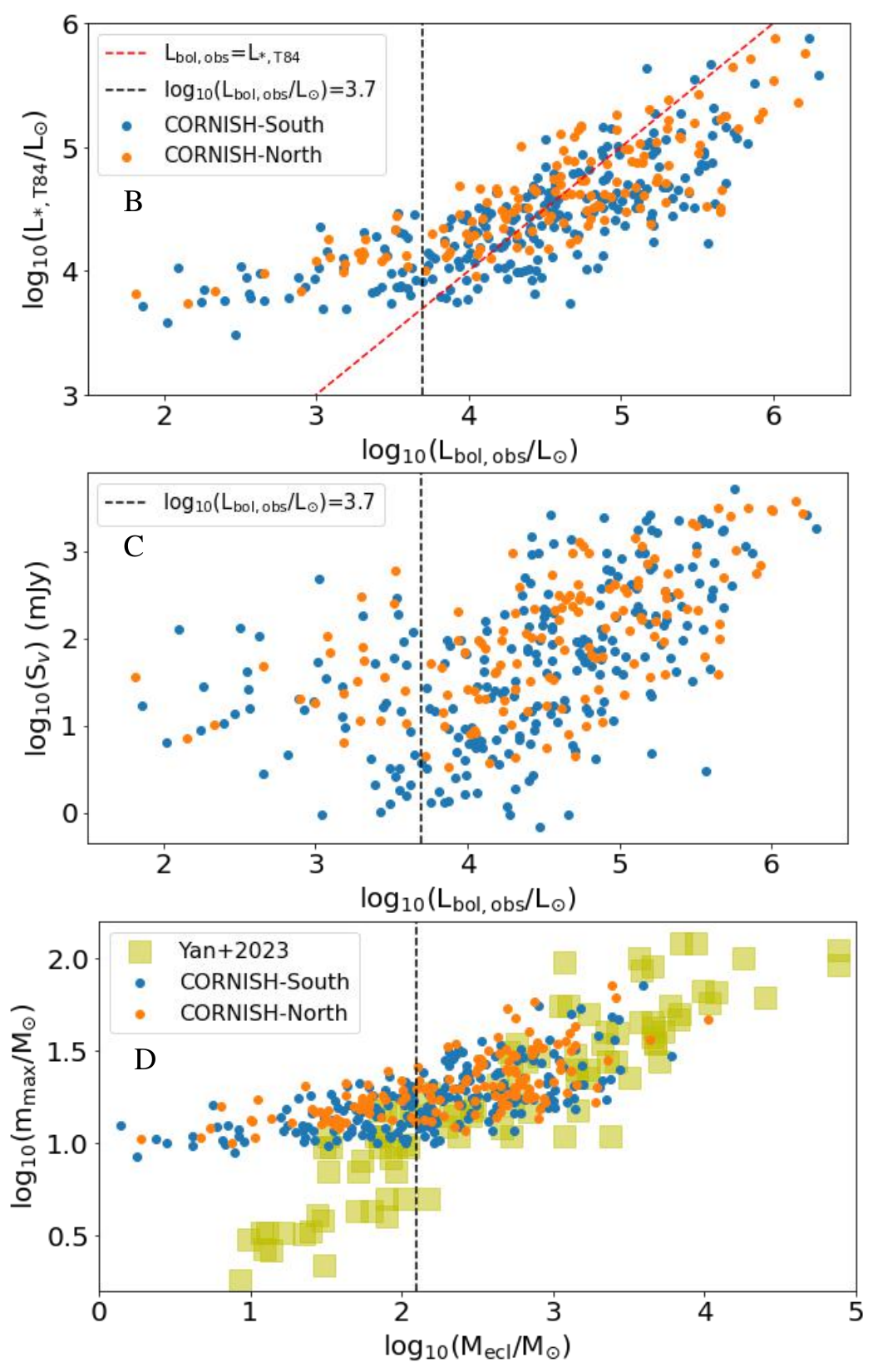}
\caption{(a) The correlation between $L_\mathrm{*, T84}$ and $L_{\rm bol, obs}$; (b) The correlation between $S_{\rm \nu}$ and $L_{\rm bol, obs}$; (c) The $m_{\rm max} -M_{\rm ecl}$ relation. Yellow squares represent the star clusters from the table of \citet{Yan2023-670}. These vertical dashed lines mark the turning points, see text for the details.}
\label{turn}
\end{figure}

The stellar luminosity $L_\mathrm{*, T84}$ of the brightest star should be smaller than the bolometric luminosity $L_{\rm bol, obs}$ of the corresponding HII-clump.
However, in Fig.~\ref{turn}(a), 
for low-luminosity HII-clumps, $L_\mathrm{*, T84}$ is significantly larger than $L_{\rm bol, obs}$, which is abnormal.
In Eq.~\ref{nl}, $N_\mathrm{Ly} \propto S_\nu$, and in Fig.~\ref{iso}, $N_\mathrm{Ly} \propto L_\mathrm{*}$. If the bolometric luminosity $L_{\rm bol, obs}$ of an HII-clump is mainly from the stars, $L_{\rm bol, obs} \propto L_\mathrm{*}$.
Finally, $L_{\rm bol, obs} \propto S_\nu$. However, in Fig.~\ref{turn}(b), there is a turning point around log$_{10}$(L$_{\rm bol, obs}$/L$_{\odot}$) $\approx$ 3.7. For HII-clumps with luminosity lower than this value, there is no clear correlation between $L_{\rm bol, obs}$ and radio flux density $S_{\rm \nu}$.
The turning point log$_{10}$(L$_{\rm bol, obs}$/L$_{\odot}$) $\approx$ 3.7 corresponds to the HII-clump mass log$_{10}$($M_{\rm cl}$/M$_{\odot}$) $\approx$ 2.55 in Fig.~\ref{sample}(a).


Using the 0.5 Myr isochrone, we estimated the stellar mass $m_\mathrm{*, iso}$ according to the stellar luminosity $L_\mathrm{*, T84}$.
The ionizing sources represent the most massive stars in HII-clumps, i.e. $m_\mathrm{*, iso} \approx m_{\rm max}$. 
Considering the $m_{\rm max}-M_{\rm ecl}$
relation \citep{Weidner2006-365,Weidner2013-434,Yan2023-670}, a correlation between the embedded star cluster mass $M_{\rm ecl}$ and the mass of the most massive star $m_{\rm max}$,
there should be a correlation between 
$m_\mathrm{*, iso}$ and 
the mass of the corresponding embedded cluster. The clump mass $M_{\rm cl}$ and its embedded cluster mass $M_{\rm ecl}$ satisfy $M_{\rm ecl}$ = SFE $\times M_{\rm cl}$, where SFE is the final star formation efficiency of the clump, and which we assumed to be SFE=0.33 \citep{Lada2003-41,Machida2012-421,Bate2014-437,Megeath2016-151,Banerjee2018-424,Zhou2024-688,Zhou2024arXiv240809867Z}, see Sec.\ref{s-sfe} for more discussion.
For 100 embedded clusters with age < 5 Myr, the table in \citet{Yan2023-670} gives the values of their $m_{\rm max}$ and $M_{\rm ecl}$. This sample fits the $m_{\rm max}-M_{\rm ecl}$ relation well.
In Fig.~\ref{turn}(c), we compared our sample with the sample from \citet{Yan2023-670}. When log$_{10}$($M_{\rm ecl}/M_{\odot}$) $>$ 2.1, the $m_\mathrm{*, iso}-M_{\rm ecl}$ relation follows the $m_{\rm max}-M_{\rm ecl}$ relation.
Assuming SFE=0.33, from the turning point
log$_{10}$($M_{\rm ecl}/M_{\odot}$) $\approx$ 2.1, we have
log$_{10}$($M_{\rm cl}/M_{\odot}$) $\approx$ 2.55, consistent with the findings above.

\subsection{Synthetic embedded clusters}
\begin{figure}
\centering
\includegraphics[width=0.47\textwidth]{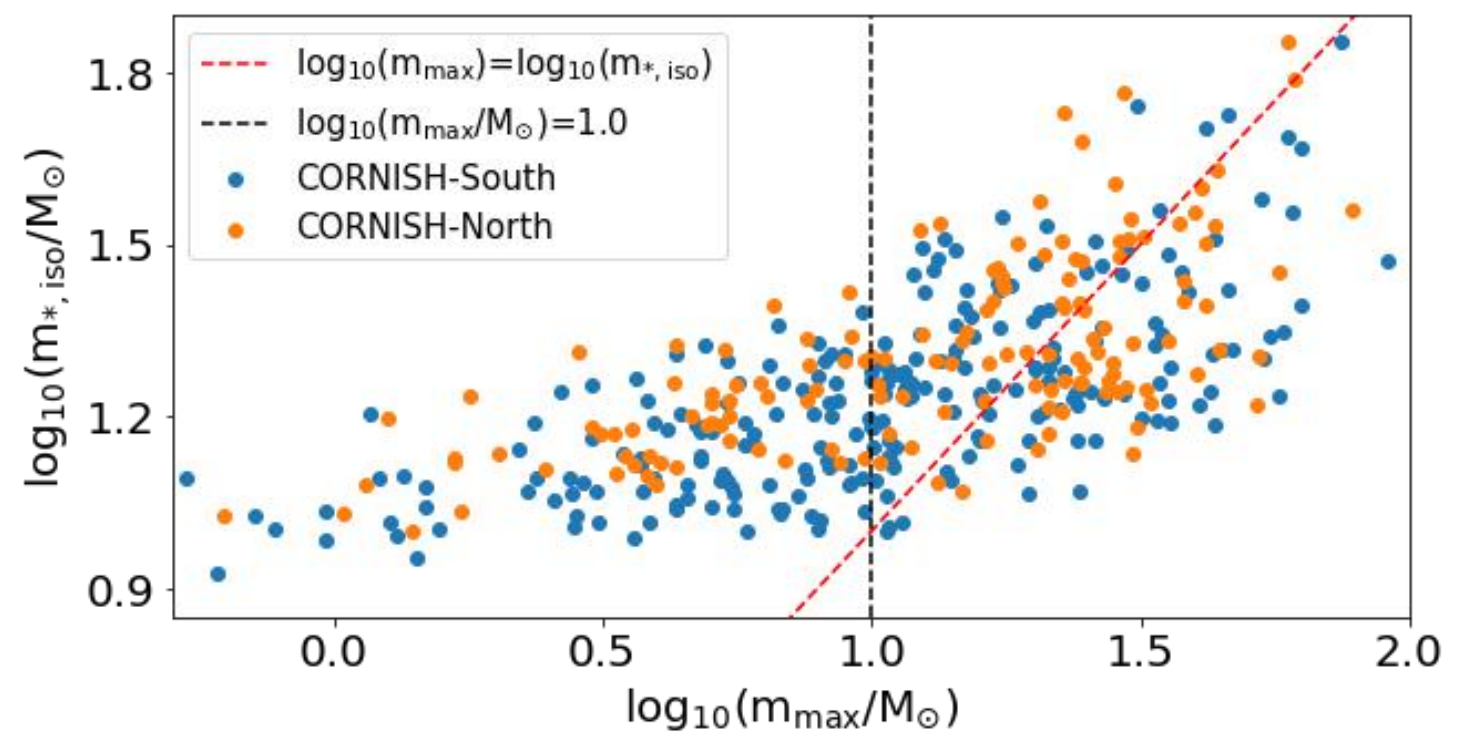}
\caption{The correlation between the most massive star with mass $m_{\rm max}$ in each synthetic embedded cluster and the corresponding $m_\mathrm{*, iso}$.}
\label{code}
\end{figure}

We adopted the optimal sampling algorithm \citep{Kroupa2013-115} to produce a population of stars for an embedded cluster with the mass $M_{\rm ecl}$ using the publicly available GalIMF code. Detailed descriptions of the optimal sampling and the publicly available GalIMF code that optimally populates clusters and galaxies with stars are available in \citet{Schulz2015-582} and \citet{Yan2017-607}. The code has three input parameters: 
the mass of the embedded cluster, $M_{\rm ecl}$, its age, and a metallicity associated with this adopted age. In this work, we assume that the metallicity of the clumps and their associated embedded clusters is the solar metallicity ([Fe/H] = 0). For the age, following \citet{Zhou2024arXiv240809867Z}, it is taken to be 1 Myr.

In Fig.~\ref{code}, we picked out the most massive star with mass $m_{\rm max}$ in each synthetic embedded cluster and compared it with the corresponding $m_\mathrm{*, iso}$. When 
log$_{10}$($m_{\rm max}/M_{\odot}$) $>$ 1, $m_{\rm max}$ and $m_\mathrm{*, iso}$ are roughly comparable. If the mass of the most massive star in a synthetic embedded cluster is log$_{10}$($m_{\rm max}/M_{\odot}$) $\approx$ 1, which corresponds to log$_{10}$($M_{\rm cl}/M_{\odot}$) $\approx$ 2.55 or log$_{10}$($M_{\rm ecl}/M_{\odot}$) $\approx$ 2.1, consistent with the results in Sec.~\ref{turning}. Moreover,
using the fitted $m_{\rm max}-M_{\rm ecl}$ relation, i.e. equation.1 in \citet{Weidner2013-434}, when log$_{10}$($M_{\rm ecl}/M_{\odot}$) $\approx$ 2.1, we can also obtain log$_{10}$($m_{\rm max}/M_{\odot}$) $\approx$ 1.

\subsection{Accretion}\label{accretion}
\begin{figure}
\centering
\includegraphics[width=0.47\textwidth]{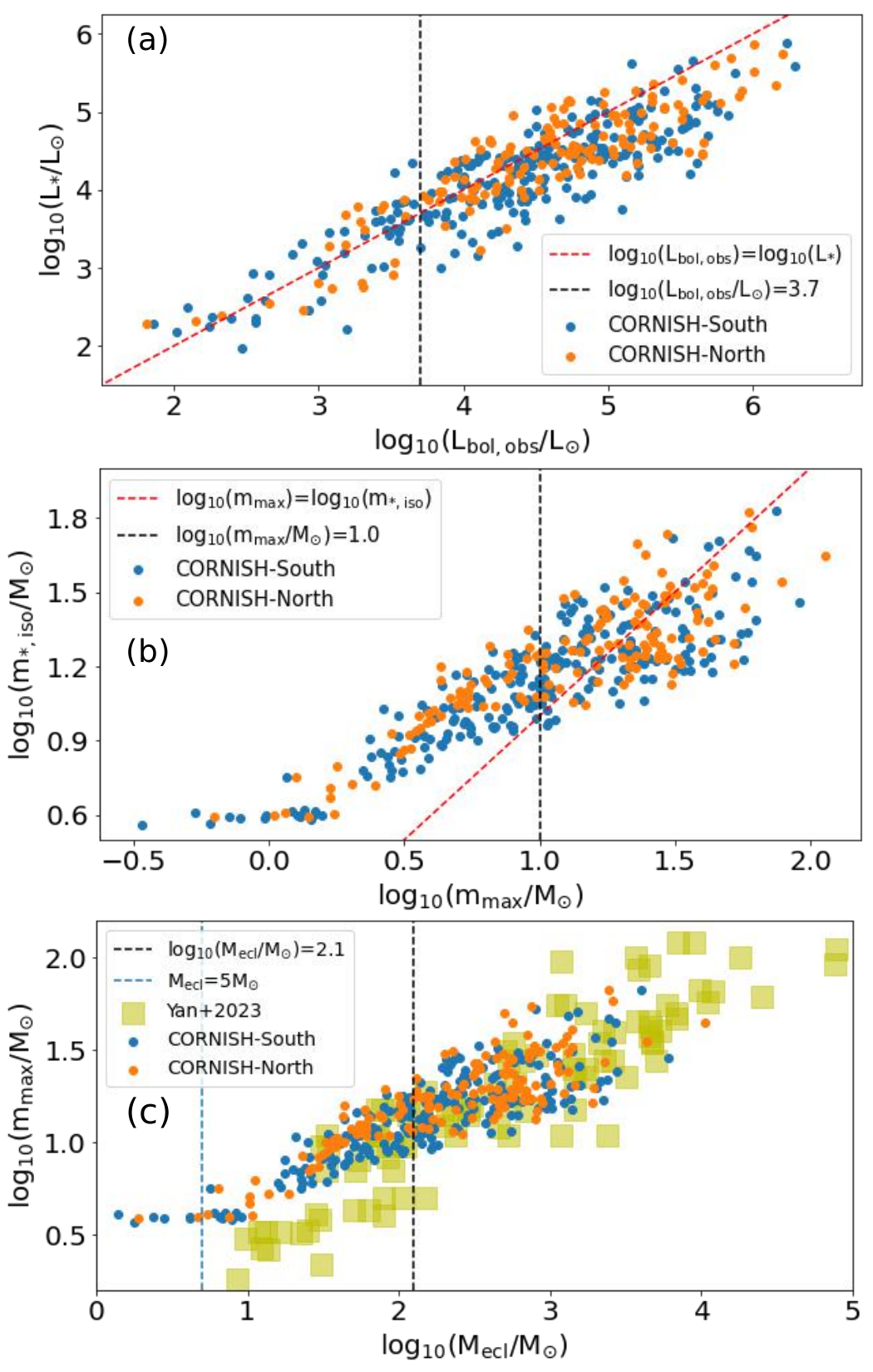}
\caption{(a) The correlation between $L_\mathrm{*}$ and $L_{\rm bol, obs}$; (b) Comparison between $m_{\rm max}$ from the synthetic clusters and $m_\mathrm{*, iso}$ from the observation; (c) The $m_{\rm max} -M_{\rm ecl}$ relation. Yellow squares represent the star clusters from the table of \citet{Yan2023-670}.}
\label{turn1}
\end{figure}

The turning point log$_{10}$(M$_{\rm cl}$/M$_{\odot}$) $\approx$ 2.55 is used to judge the relative importance of accretion luminosity in \citet{Zhou2024arXiv240809867Z}, also presented in Fig.~\ref{sample}(a). 
Therefore, the abnormal tail in Fig.~\ref{turn} and Fig.~\ref{code} on the left of the vertical dashed lines means that if the accretion luminosity is dominant, the formalism in Sec.~\ref{Lyman} cannot be used, because the main Lyman continuum flux or radio emission are not predominantly from the stars.
The high radio flux density of low-luminosity and low-mass HII-clumps in Fig.~\ref{turn}(b) indicate that accretion is an effective way to produce radio emission. Therefore, the Lyman continuum flux $N_\mathrm{Ly}$ derived from radio flux density $S_{\rm \nu}$ includes a significant contribution from the accretion process. The previous studies of HII regions in \citet{Lumsden2013-208, Urquhart2013-435, Sanchez-Monge2013-550} have noted that there is an excess of Lyman photons over what would be expected from the ionizing stars compared to the bolometric luminosities. The inclusion of the luminosity contribution from accretion may provide an explanation.

In Sec.~\ref{Lyman}, we obtained the stellar luminosity $L_\mathrm{*, T84}$ from the Lyman continuum flux $N_\mathrm{Ly}$. Actually, $L_\mathrm{*, T84}$ is not only the stellar luminosity, which includes the real stellar luminosity $L_\mathrm{*}$, but the sum of $L_\mathrm{*}$ and the accretion luminosity $L_\mathrm{acc}$, i.e. $L_\mathrm{*, T84}$=$L_\mathrm{*}$+$L_\mathrm{acc}$. 
Using the method described in \citet{Zhou2024arXiv240809867Z}, we calculated the accretion luminosity $L_\mathrm{acc}$ of the most massive stars of the synthetic embedded clusters, and subtracted $L_\mathrm{acc}$ from the corresponding $L_\mathrm{*, T84}$. 
Then we repeated the analysis in Sec.~\ref{turning} using $L_\mathrm{*}$ rather than $L_\mathrm{*, T84}$.
As can be seen in Fig.~\ref{turn1}(a), the abnormal tail disappears.

\subsection{Flares}\label{s-flare}

\begin{figure}
\centering
\includegraphics[width=0.47\textwidth]{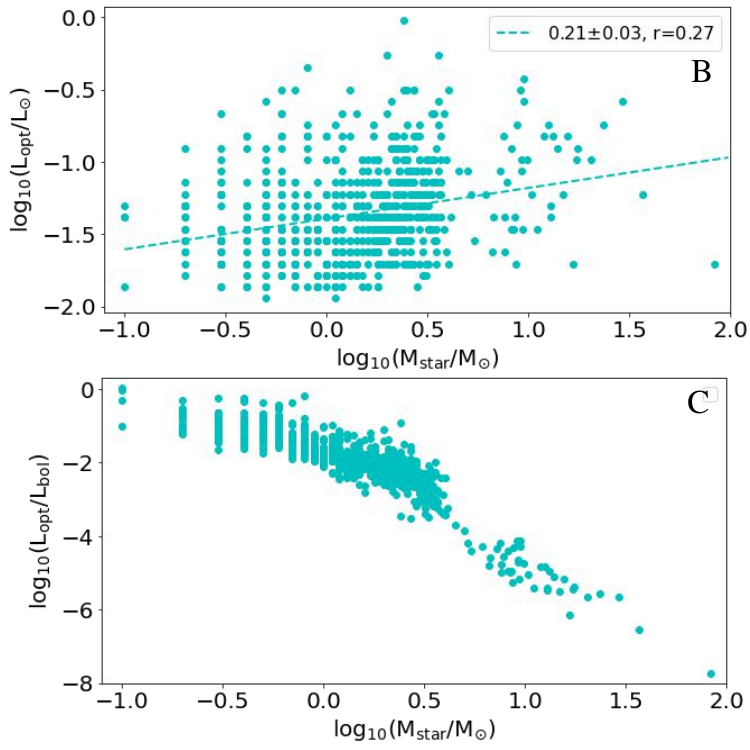}
\caption{(a) Correlation between the optical luminosity and the stellar mass. "r" is the Pearson correlation coefficient of the linear fitting; (b) Correlation between the ratio of optical and bolometric luminosities of a flare and the stellar mass.}
\label{flare}
\end{figure}

The value of $L_\mathrm{*}$ still remains somewhat higher in Fig.~\ref{turn1}(a). The abnormal tail in Fig.~\ref{turn1}(b) is not completely eliminated, although it shows a substantial improvement over the case without considering accretion, presented in Fig.~\ref{code}. 
Therefore,
the stellar luminosity $L_\mathrm{*}$ may still be overestimated. There are likely other physical processes contributing to the luminosity that have not been considered, similar to the accretion luminosity, which also needs to be subtracted.

Flaring represents a common occurrence of magnetic activity in low-mass stars, including our Sun. While flares, originating from both the Sun and other stars, are predominantly observed in the soft X-ray band, the majority of the emitted energy is released at optical/UV wavelengths \citep{Flaccomio2018-620,Kuznetsov2021-912}.
For a  sample of pre-main-sequence (PMS) stars in the NGC 2264 star-forming region, \citet{Flaccomio2018-620} detected seventy-eight X-ray flares with optical and/or mid-infrared counterparts. 
The optical emission of flares, encompassing both emitted energy and peak flux, demonstrates a strong correlation with, and notably surpasses, the X-ray emission.
The luminosities in X-ray ($L_{\rm x}$) and optical ($L_{\rm opt}$) bands satisfy $L_{\rm opt}= a \times L_{\rm x}^{b}$ with $a$ ranging from 5.2 to 14.7 and $b$ between 0.6 and 1.0. Here we take the average values, i.e. $L_{\rm opt} \approx 10 \times L_{\rm x}^{0.8}$. 

\citet{Getman2021-916} identified a sample of 1086 X-ray superflares and megaflares.
These events are generated by young stars across all masses throughout evolutionary stages spanning from protostars to stars without disks.
Only the sample of \citet{Getman2021-916} has mass estimates of the PMS stars. We converted the X-ray luminosity of the flare measured in \citet{Getman2021-916} into optical luminosity using the relation fitted in \citet{Flaccomio2018-620}. In Fig.~\ref{flare}(a), the distribution of $L_{\rm opt}$ is highly scattered, and it has almost no correlation with stellar mass. The ratio of the optical and bolometric luminosities ($L_{\rm opt}/L_{\rm bol}$) of the PMS stars decreases rapidly as mass increases. Even for low-mass stars, the ratio is always less than 1. Thus, the luminosity from the flares can be neglected, and does not make a significant contribution to the abnormal tail in Sec.~\ref{turning}.


\subsection{Star formation efficiency}\label{s-sfe}
\begin{figure*}
\centering
\includegraphics[width=0.95\textwidth]{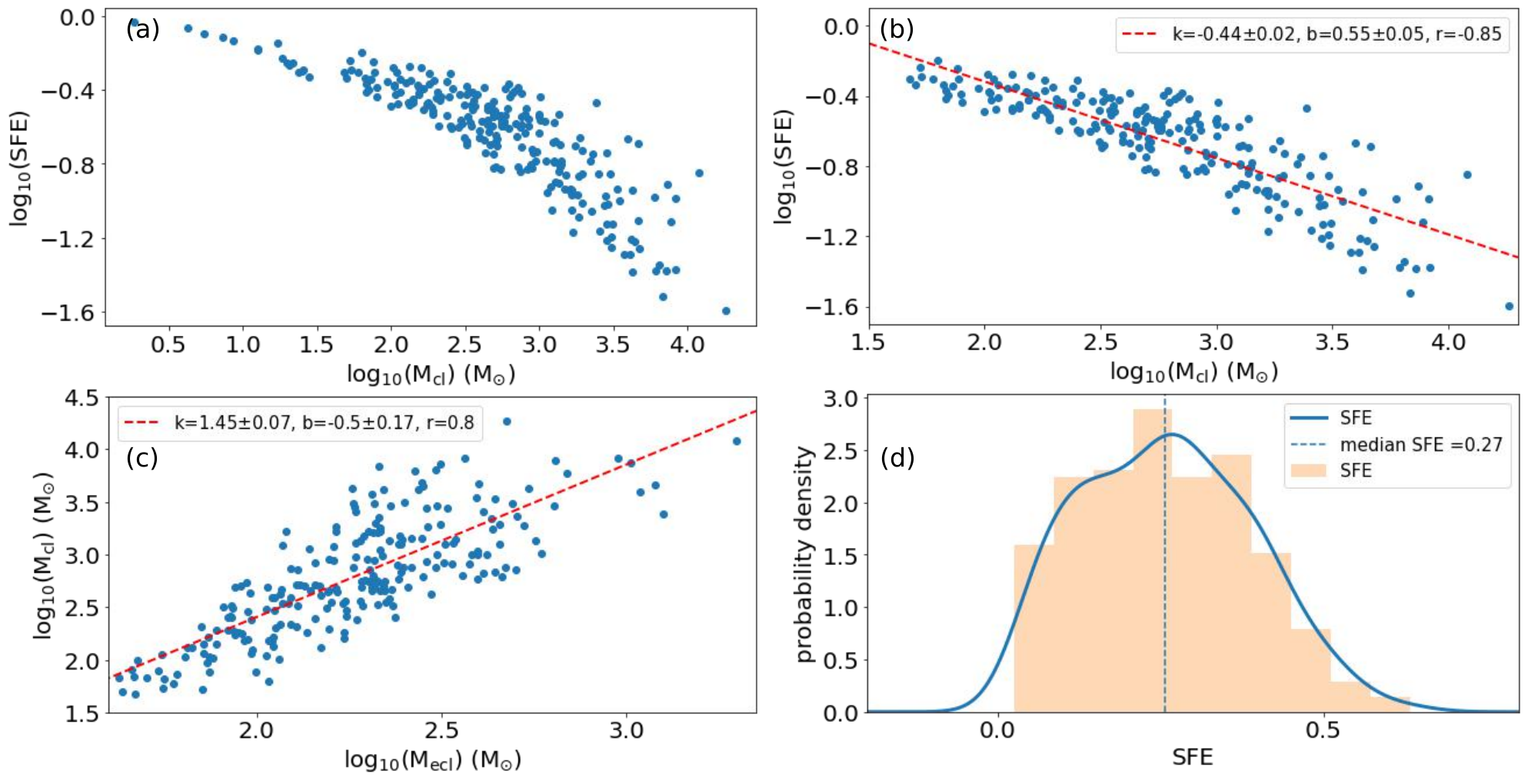}
\caption{Star formation efficiency derived from the observations.
(a) The correlation between the star formation efficiency and the clump mass for all the data points presented in Fig.\ref{turn1}(b); (b) Eliminated the abnormal tail in panel (a); (c) The correlation between the embedded cluster mass and the corresponding HII-clump mass; (d) The distribution of the SFEs. Red dashed line is the linear fitting to the data points. "k" and "b" are the slope and the intercept, "r" is the Pearson correlation coefficient.}
\label{sfe-o}
\end{figure*}

\begin{figure}
\centering
\includegraphics[width=0.475\textwidth]{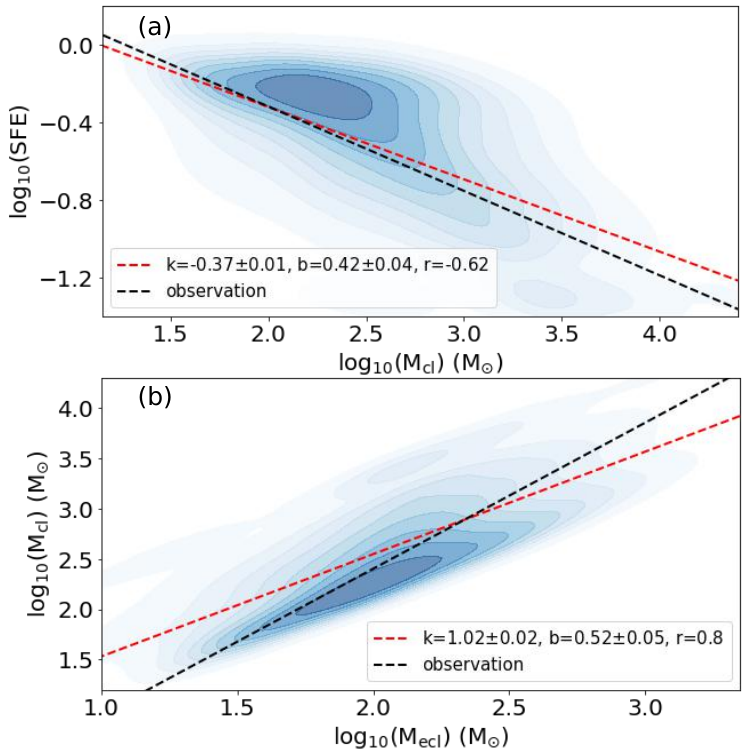}
\caption{Star formation efficiency derived from the model.
(a) The correlation between the star formation efficiency and the clump mass for all HII-clumps with a mass larger than 10$^{1.5}$ M$_{\odot}$; (b) The correlation between the embedded cluster mass and the corresponding HII-clump mass. Red dashed line is the linear fitting to the data points. Black dashed lines come from Fig.\ref{sfe-o}.
"k" and "b" are the slope and the intercept, "r" is the Pearson correlation coefficient.}
\label{sfe-m}
\end{figure}

\begin{figure}
\centering
\includegraphics[width=0.475\textwidth]{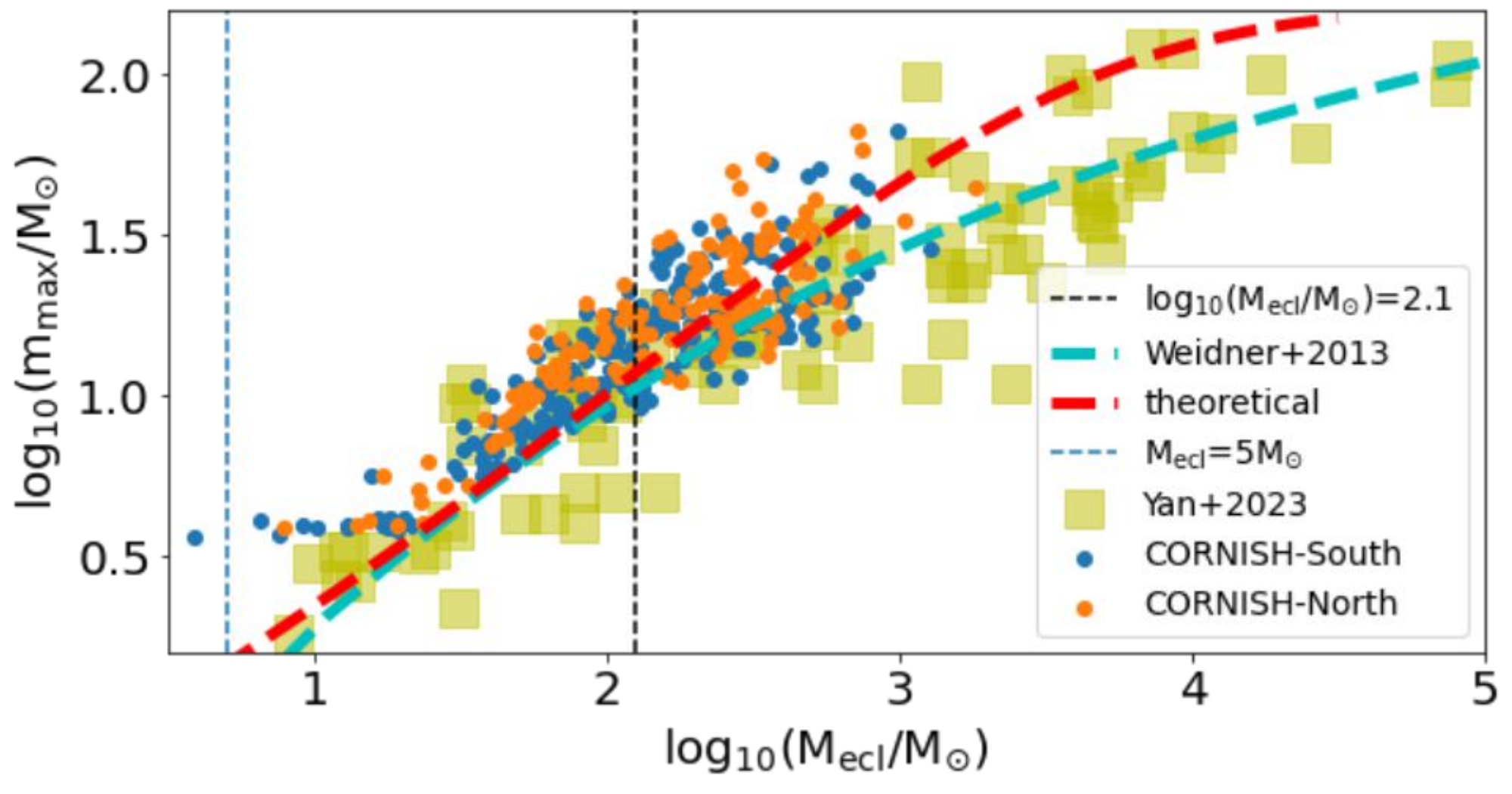}
\caption{Same as Fig.\ref{turn1}(c), but after modifying the SFE using equation.\ref{sfe-mcl}. Cyan dashed line is the $m_{\rm max}-M_{\rm ecl}$ relation fitted in \citet{Weidner2013-434} based on the observation. Red dashed line is the $m_{\rm max}-M_{\rm ecl}$ relation derived from the IMF theory, i.e. equation.\ref{mmaxM}.}
\label{mod}
\end{figure}

\begin{figure}
\centering
\includegraphics[width=0.475\textwidth]{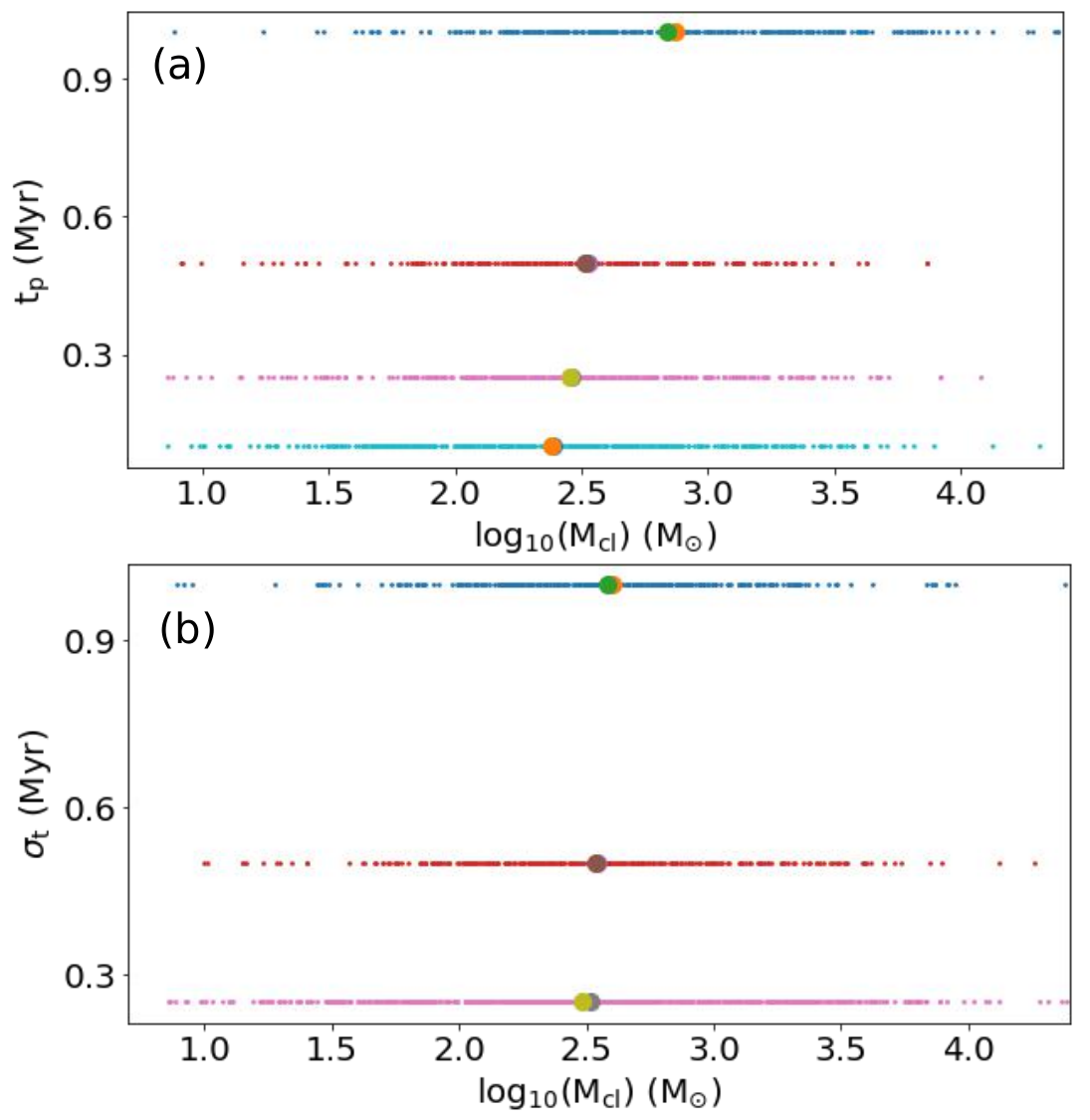}
\caption{The distributions of the optimal $t_{\rm p}$ and $\sigma_{\rm t}$ for all HII-clumps with mass estimation. For each $t_{\rm p}$ or $\sigma_{\rm t}$, we calculated the median and mean masses of the corresponding HII-clump groups, marked by big dots.}
\label{sfh}
\end{figure}

\subsubsection{Observations}

In Fig.~\ref{turn1}(c), for low-mass clusters, there is a mismatch between our sample and the sample in \citet{Yan2023-670}. As discussed in Sec.~\ref{s-flare}, we may be overestimating the stellar mass. 
Another possibility is that the low-mass clumps have higher SFEs > 0.33. Assuming all measured maximum stellar masses follow the $m_{\rm max}-M_{\rm ecl}$ relation, we can calculate the mass of the embedded cluster and then the star formation efficiency for each clump.
The third-order polynomial fit to the observed $m_{\rm max}-M_{\rm ecl}$ relation presented in \citet{Weidner2013-434} has large errors.
As an alternative, we derived a theoretical $m_{\rm max}-M_{\rm ecl}$ relation directly from the initial mass function (IMF). The details are described in Sec.\ref{mM} and the embedded cluster mass, $M_{\rm ecl}$, is given by

\begin{equation}
\begin{split}
& M_{\mathrm{ecl}} \approx 
\frac{5.37}{\frac{0.77}{m_{\mathrm{max}}^{1.3}} - 0.001} - \frac{3.33}{m_{\mathrm{max}}^{0.3} \left(\frac{0.77}{m_{\mathrm{max}}^{1.3}} - 0.001\right)},
\end{split}
\label{mmaxM}
\end{equation}

\noindent where we use the $m_{\rm max}$ values displayed in Fig~.\ref{turn1}(c) to derive the corresponding value of $M_{\rm ecl}$ for each embedded cluster.

The abnormal tail in Fig.~\ref{turn1}(b) is visible in Fig.~\ref{sfe-o}(a) again, presented as the SFEs close to 1. We excluded this tail in the subsequent analysis by requiring the clump mass larger than 10$^{1.5}$ M$_{\odot}$.
Fig.~\ref{sfe-o}(b) displays a strong correlation between the SFE and the clump mass. The SFE decreases with increasing clump mass, with an upper limit of $\approx$0.65, This is also visible in Fig.5 of \citet{Wells2022-516}. We fit the relation between $M_{\rm cl}$ and the SFE with a power law of the form ${\rm SFE} \propto M_{\rm cl}^{k}$ and find a value of $k=-0.44\pm0.02$. 
This anti-correlation between the mass of the clumps and their SFE has been predicted theoretically by \citet{Dib2013-436}. For the simple case of uniform density clumps, \citet{Dib2013-436} predicted the clump mass and the SFE to follow ${\rm SFE} \propto M_{\rm cl}^{-0.6}$ and to be the result of an increasingly stronger effect of stellar feedback in clumps with higher masses which leads to a faster gas expulsion and limits the value of the SFE. Interestingly, the median value of the SFEs is $\approx$0.3, consistent with our previous work \citep{Zhou2024-688,Zhou2024arXiv240809867Z}. An SFE$\approx 0.33$ is also consistent with the value obtained from hydrodynamic calculations including self-regulation \citep{Machida2012-421,Bate2014-437} and also with observations of embedded systems in the solar neighborhood \citep{Lada2003-41,Megeath2016-151}.


\subsubsection{Models}

In order to assess the uncertainty in the observations, now we independently study the clump's SFE using theoretical models. In Fig.~\ref{sfe-o}(b), the upper limit of the SFE is $\approx$0.65. In Fig. 4 of \citet{Wells2022-516}, a SFE larger than 0.5 is necessary to fit the low-mass HII-clumps.
Therefore, we consider the SFEs that fall in the range [0.05, 0.65] with a step of 0.05. Then each HII-clump has 13 alternative SFEs. 
We consider both the stellar luminosity and accretion luminosity of each star in the embedded cluster of a HII-clump, and also apply the pre-main sequence evolutionary track for each star. 
There is certainly an age spread of the stellar populations in embedded clusters within HII-clumps. Therefore, selecting the appropriate isochrones according to the stars' ages is essential. Additionally, not all stars in HII-clumps are in the accretion phase, as this too is age-dependent. We should identify the fraction of stellar objects that remain protostars to properly add their accretion luminosities. In order to address these issues, we apply the method described in \citet{Zhou2024arXiv240809867Z} to generate an age distribution for the stellar population in each synthetic cluster. The age distribution of stars within a HII-clump depends on the star formation history (SFH) of the clump. As discussed in \citet{Zhou2024arXiv240809867Z}, compared to a constant
SFH, burst-like and time-dependent SFHs can better fit the observational data. Thus, we only consider the time-dependent SFHs in this work. As described in \citet{Zhou2024arXiv240809867Z}, there are 12 different SFHs with different age peaks ($t_{\rm p}$) and age standard deviations ($\sigma_{\rm t}$), i.e. $t_{\rm p}= 0.1, 0.25, 0.5, 1$ Myr and $\sigma_{\rm t}=$ 0.25, 0.5, 1 Myr. Finally, each clump has 156 different combinations of SFEs and SFHs, resulting in 156 distinct bolometric luminosities. The combination of a SFE and a SFH that produces the bolometric luminosity of the synthetic embedded cluster closest to the observed bolometric luminosity of the corresponding HII-clump is considered as the optimal one. The optimal SFE and SFH distributions of all HII-clumps are shown in Fig.~\ref{sfe-m}. The optimal combination has the smallest difference from the observed bolometric luminosity.
To test the robustness of the model, we also selected the second optimal combination, which has the second smallest difference from the observed bolometric luminosity. As shown in Fig.~\ref{sfe-ms}, the optimal and the second optimal combinations give almost the same results.

Compared with Fig.~\ref{sfe-o}, overall, the SFEs of HII-clumps derived from the observation and the model are comparable. 
From Fig.~\ref{sfe-m}(b) and (c), we have

\begin{equation}
\mathrm{log}_{10} (\mathrm{SFE}) = (-0.37 \pm 0.01) \times \mathrm{log}_{10} (M_{\mathrm{cl}}) + (0.42 \pm 0.04),
\label{sfe-mcl}
\end{equation}
and
\begin{equation}
\mathrm{log}_{10} (M_{\mathrm{cl}}) = (1.02 \pm 0.02) \times \mathrm{log}_{10} (M_{\mathrm{ecl}}) + (0.52 \pm 0.05).
\end{equation}
Then we use Eq.~\ref{sfe-mcl} to calculate the SFE or the mass of the embedded cluster for each HII-clump. As shown in Fig.~\ref{mod},
after the modification of the SFE, the mismatch in Fig.~\ref{turn1}(c) indeed disappears.

\subsection{Star formation history}

\begin{figure}
\centering
\includegraphics[width=0.475\textwidth]{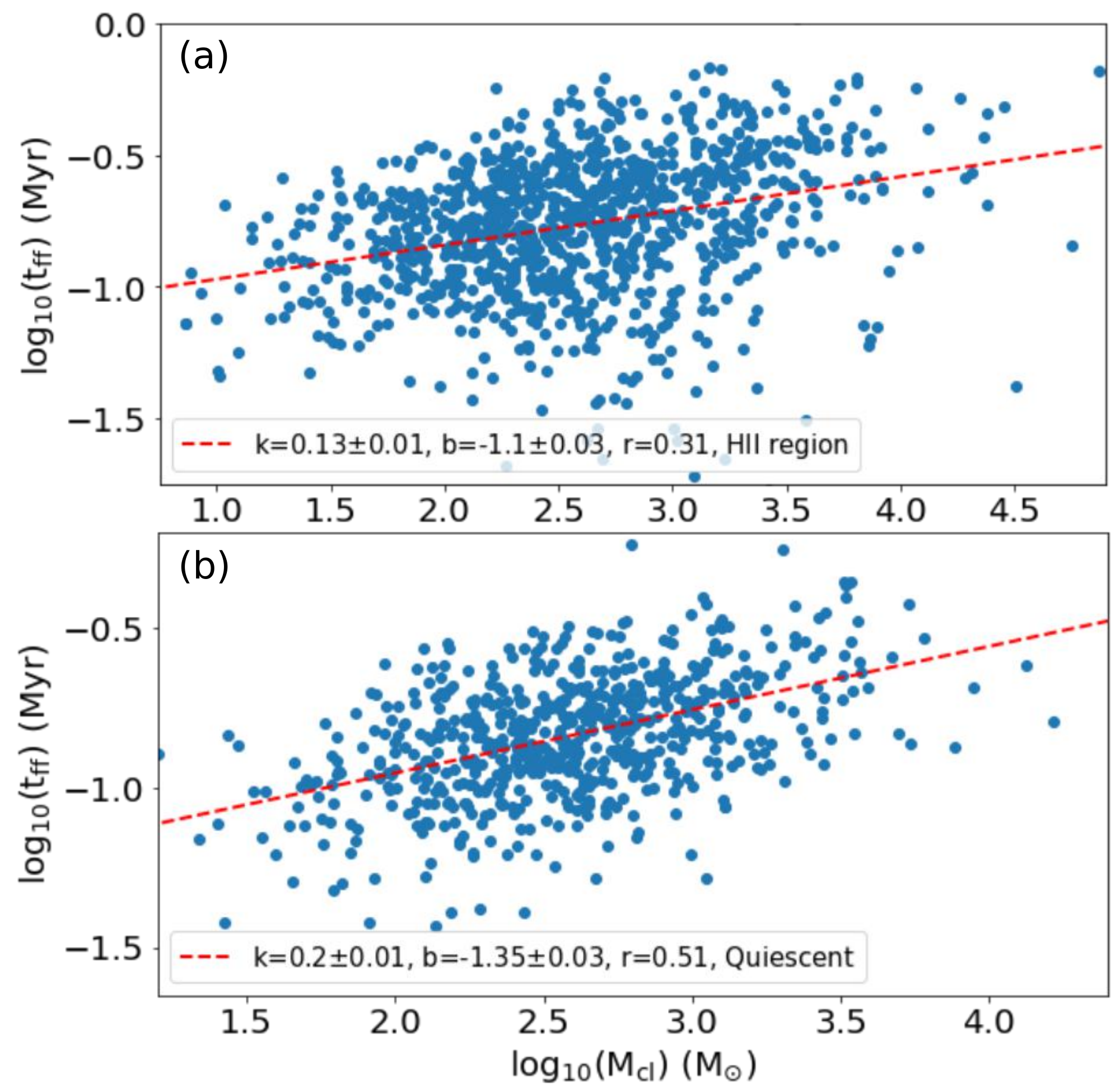}
\caption{The correlation between the free-fall time and the clump mass in the earliest ("quiescent") and the latest ("HII region") stages.}
\label{tff}
\end{figure}

Fig.~\ref{sfh} displays the distributions of the optimal $t_{\rm p}$ and $\sigma_{\rm t}$ for HII-clumps with different masses. For each $t_{\rm p}$ or $\sigma_{\rm t}$, we calculated the median and mean masses of the corresponding HII-clump groups. The median and mean clump masses are comparable for different $\sigma_{\rm t}$. However, there is a weak trend that clumps with larger masses have larger $t_{\rm p}$, indicating that low-mass clumps reach the peak of star formation earlier than high-mass clumps, possibly because star formation occurs more rapidly in low-mass clumps. The clump free-fall time ($\tau_{\rm ff}$) is useful in providing a lower limit to the star formation time-scales. In Fig.\ref{tff}, we compared the free-fall time of ATLASGAL clumps in the earliest ("quiescent") and the latest ("HII region") stages \citep{Urquhart2022-510}. 
Low-mass clumps indeed have a shorter free-fall time.


\section{Conclusion}

In this work, 1226 HII-clumps from the ATLASGAL survey with mass $M_{\rm cl}$ and bolometric luminosity $L_{\rm bol, obs}$ estimates were matched with radio sources in the catalogs of the CORNISH-North and the CORNISH-South
surveys. 392 HII-clumps have corresponding radio sources in the CORNISH-North/South surveys. These HII-clumps are mainly concentrated at the
upper end in the $L_{\rm bol, obs}$-$M_{\rm cl}$ diagram, which are more luminous than the remaining HII-clumps, indicating they host high-mass stars. Radio sources are excited by the most massive stars in the embedded clusters of HII-clumps. We calculated the emission measure ($EM$), the electron density ($n_\mathrm{e}$) and the Lyman continuum flux ($N_\mathrm{Ly}$) for each radio source. Using  the relationship between the stellar parameters and the Lyman continuum flux in Table. 1 of \citet{Thompson1984-283}, 
we determined the stellar luminosity $L_\mathrm{*, T84}$ according to the calculated $N_\mathrm{Ly}$. 
We found a turning point around log$_{10}$(L$_{\rm bol, obs}$/L$_{\odot}$) $\approx$ 3.7, which corresponds to a HII-clump mass log$_{10}$(M$_{\rm cl}$/M$_{\odot}$) $\approx$ 2.55. When the bolometric luminosity of HII-clumps is less than this value, $L_\mathrm{*, T84}$ is significantly larger than $L_{\rm bol, obs}$. Thus, there is an abnormal tail in the $L_{\rm bol, obs}$-$L_\mathrm{*, T84}$ diagram. Using the 0.5 Myr isochrone, we estimated the stellar mass $m_\mathrm{*, iso}$ according to the stellar luminosity $L_\mathrm{*, T84}$.

To investigate the physical origin of this turning point, we adopted the optimal sampling algorithm to produce a population of stars for an embedded cluster using the publicly available GalIMF code.
Then, we picked out the most massive star with mass $m_{\rm max}$
in each synthetic embedded cluster and compared it with the corresponding $m_\mathrm{*, iso}$. When log$_{10}$($m_{\rm max}/M_{\odot}$) > 1, $m_{\rm max}$ and $m_\mathrm{*, iso}$ are roughly comparable.
Actually, the turning point at log$_{10}$(M$_{\rm cl}$/M$_{\odot}$) $\approx$ 2.55 exactly corresponds to log$_{10}$($m_{\rm max}/M_{\odot}$) $\approx$ 1, which is also used to judge the relative importance of accretion luminosity in \citet{Zhou2024arXiv240809867Z}. 
The abnormal tail means that if the accretion luminosity is dominant, we can not directly estimate the stellar parameters from the radio emission, and have to subtract the contribution of accretion from the radio flux density. To address this issue, we calculated the accretion luminosity $L_\mathrm{acc}$ of the most massive star in each synthetic embedded cluster, and subtracted $L_\mathrm{acc}$ from the corresponding $L_\mathrm{*, T84}$.
However, the cleaned stellar luminosity $L_\mathrm{*}$ (=$L_\mathrm{*, T84}$-$L_\mathrm{acc}$) may be still overestimated. There are likely other physical processes contributing to the luminosity apart from the accretion. For example, the flaring activity represents a common occurrence of magnetic activity in low-mass stars. We found that the ratio of the optical and bolometric luminosities ($L_{\rm opt}/L_{\rm bol}$) of pre-main sequence stars decreases rapidly as mass increases. Even for low-mass stars, the ratio is always less than
1. Thus, the luminosity from the flaring activity can be neglected. Other physical processes need to be further investigated.

In the above analysis, we assumed a constant SFE=0.33. We further studied the change of SFE with the clump mass. According to the derived mass of the most massive star in each HII-clump, using the theoretical $m_{\rm max}-M_{\rm ecl}$ relation, we calculated the mass of the corresponding embedded cluster and then the SFE of the clump. We find a strong anti-correlation between the SFE and the clump mass. The SFE decreases with increasing clump mass in agreement with theoretical expectations \citep{Dib2013-436}, with a median value of $\approx$0.3.   
We also independently derived the SFE for each HII-clump based on the model developed in \citet{Zhou2024arXiv240809867Z}. We consider the SFEs that fall in the range [0.05, 0.65] with a step of 0.05. 
We employ 12 different star formation histories for each HII-clump to distribute the ages to the stars in the clump. According to the stars' age,
we consider both the stellar luminosity and accretion luminosity of each star in the embedded cluster of a HII-clump, and apply the pre-main sequence evolutionary track for each star. Finally,
each clump has 156 different combinations of SFEs and SFHs, resulting in 156 distinct bolometric luminosities. The optimal one has the smallest difference from the observed bolometric luminosity of the corresponding HII-clump. Overall, the SFEs of HII-clumps derived from the observations and the models are comparable. 
Using the quantitative relation between the SFE and the clump mass, we calculated the SFE and the mass of the embedded cluster for each HII-clump. As shown in Fig.~\ref{mod},
after the modification of the SFE, the mismatch in Fig.~\ref{turn1}(c) disappears.

For the optimal $t_{\rm p}$ or $\sigma_{\rm t}$, we estimated the median and mean masses of the corresponding HII-clump groups. There is a weak trend that clumps with larger masses have larger $t_{\rm p}$, indicating that low-mass clumps reach the peak of star formation earlier than high-mass clumps, possibly because star formation occurs more rapidly in low-mass clumps, consistent with the shorter free-fall time of low-mass clumps.

\begin{acknowledgements}
We would like to thank the referee for the comments and suggestions that helped improve and clarify this work.
\end{acknowledgements}

\bibliographystyle{aa} 
\bibliography{ref}

\begin{thebibliography}{50}
\expandafter\ifx\csname natexlab\endcsname\relax\def\natexlab#1{#1}\fi

\bibitem[{{Banerjee} \& {Kroupa}(2018)}]{Banerjee2018-424}
{Banerjee}, S. \& {Kroupa}, P. 2018, in Astrophysics and Space Science Library, Vol. 424, The Birth of Star Clusters, ed. S.~{Stahler}, 143

\bibitem[{{Bate} {et~al.}(2014){Bate}, {Tricco}, \& {Price}}]{Bate2014-437}
{Bate}, M.~R., {Tricco}, T.~S., \& {Price}, D.~J. 2014, \mnras, 437, 77

\bibitem[{{Choi} {et~al.}(2016){Choi}, {Dotter}, {Conroy}, {Cantiello}, {Paxton}, \& {Johnson}}]{Choi2016-823}
{Choi}, J., {Dotter}, A., {Conroy}, C., {et~al.} 2016, \apj, 823, 102

\bibitem[{{Dabringhausen} {et~al.}(2009){Dabringhausen}, {Kroupa}, \& {Baumgardt}}]{Dabringhausen2009-394}
{Dabringhausen}, J., {Kroupa}, P., \& {Baumgardt}, H. 2009, \mnras, 394, 1529

\bibitem[{{Davies} {et~al.}(2011){Davies}, {Hoare}, {Lumsden}, {Hosokawa}, {Oudmaijer}, {Urquhart}, {Mottram}, \& {Stead}}]{Davies2011-416}
{Davies}, B., {Hoare}, M.~G., {Lumsden}, S.~L., {et~al.} 2011, \mnras, 416, 972

\bibitem[{{Dib} {et~al.}(2013){Dib}, {Gutkin}, {Brandner}, \& {Basu}}]{Dib2013-436}
{Dib}, S., {Gutkin}, J., {Brandner}, W., \& {Basu}, S. 2013, \mnras, 436, 3727

\bibitem[{{Dib} {et~al.}(2007){Dib}, {Kim}, {V{\'a}zquez-Semadeni}, {Burkert}, \& {Shadmehri}}]{Dib2007-661}
{Dib}, S., {Kim}, J., {V{\'a}zquez-Semadeni}, E., {Burkert}, A., \& {Shadmehri}, M. 2007, \apj, 661, 262

\bibitem[{{Dinnbier} {et~al.}(2022){Dinnbier}, {Kroupa}, \& {Anderson}}]{Dinnbier2022-660}
{Dinnbier}, F., {Kroupa}, P., \& {Anderson}, R.~I. 2022, \aap, 660, A61

\bibitem[{{Elia} {et~al.}(2021){Elia}, {Merello}, {Molinari}, {Schisano}, {Zavagno}, {Russeil}, {M{\`e}ge}, {Martin}, {Olmi}, {Pestalozzi}, {Plume}, {Ragan}, {Benedettini}, {Eden}, {Moore}, {Noriega-Crespo}, {Paladini}, {Palmeirim}, {Pezzuto}, {Pilbratt}, {Rygl}, {Schilke}, {Strafella}, {Tan}, {Traficante}, {Baldeschi}, {Bally}, {di Giorgio}, {Fiorellino}, {Liu}, {Piazzo}, \& {Polychroni}}]{Elia2021-504}
{Elia}, D., {Merello}, M., {Molinari}, S., {et~al.} 2021, \mnras, 504, 2742

\bibitem[{{Elmegreen} \& {Shadmehri}(2003)}]{Elmegreen2003-338}
{Elmegreen}, B.~G. \& {Shadmehri}, M. 2003, \mnras, 338, 817

\bibitem[{{Figer}(2005)}]{Figer2005-434}
{Figer}, D.~F. 2005, \nat, 434, 192

\bibitem[{{Flaccomio} {et~al.}(2018){Flaccomio}, {Micela}, {Sciortino}, {Cody}, {Guarcello}, {Morales-Calder{\`o}n}, {Rebull}, \& {Stauffer}}]{Flaccomio2018-620}
{Flaccomio}, E., {Micela}, G., {Sciortino}, S., {et~al.} 2018, \aap, 620, A55

\bibitem[{{Getman} \& {Feigelson}(2021)}]{Getman2021-916}
{Getman}, K.~V. \& {Feigelson}, E.~D. 2021, \apj, 916, 32

\bibitem[{{Irabor} {et~al.}(2023){Irabor}, {Hoare}, {Burton}, {Cotton}, {Diamond}, {Dougherty}, {Ellingsen}, {Fender}, {Fuller}, {Garrington}, {Goldsmith}, {Green}, {Gunn}, {Jackson}, {Kurtz}, {Lumsden}, {Marti}, {McDonald}, {Molinari}, {Moore}, {Mutale}, {Muxlow}, {O'Brien}, {Oudmaijer}, {Paladini}, {Pandian}, {Paredes}, {Richards}, {Sanchez-Monge}, {Spencer}, {Thompson}, {Umana}, {Urquhart}, {Wieringa}, \& {Zijlstra}}]{Irabor2023-520}
{Irabor}, T., {Hoare}, M.~G., {Burton}, M., {et~al.} 2023, \mnras, 520, 1073

\bibitem[{{Kalcheva} {et~al.}(2018){Kalcheva}, {Hoare}, {Urquhart}, {Kurtz}, {Lumsden}, {Purcell}, \& {Zijlstra}}]{Kalcheva2018-615}
{Kalcheva}, I.~E., {Hoare}, M.~G., {Urquhart}, J.~S., {et~al.} 2018, \aap, 615, A103

\bibitem[{{Koen}(2006)}]{Koen2006-365}
{Koen}, C. 2006, \mnras, 365, 590

\bibitem[{{Kroupa}(1995{\natexlab{a}})}]{Kroupa1995a-277}
{Kroupa}, P. 1995{\natexlab{a}}, \mnras, 277, 1491

\bibitem[{{Kroupa}(1995{\natexlab{b}})}]{Kroupa1995b-277}
{Kroupa}, P. 1995{\natexlab{b}}, \mnras, 277, 1507

\bibitem[{{Kroupa}(2001)}]{Kroupa2001-322}
{Kroupa}, P. 2001, \mnras, 322, 231

\bibitem[{{Kroupa}(2005)}]{Kroupa2005-576}
{Kroupa}, P. 2005, in ESA Special Publication, Vol. 576, The Three-Dimensional Universe with Gaia, ed. C.~{Turon}, K.~S. {O'Flaherty}, \& M.~A.~C. {Perryman}, 629

\bibitem[{{Kroupa} {et~al.}(2013){Kroupa}, {Weidner}, {Pflamm-Altenburg}, {Thies}, {Dabringhausen}, {Marks}, \& {Maschberger}}]{Kroupa2013-115}
{Kroupa}, P., {Weidner}, C., {Pflamm-Altenburg}, J., {et~al.} 2013, in Planets, Stars and Stellar Systems. Volume 5: Galactic Structure and Stellar Populations, ed. T.~D. {Oswalt} \& G.~{Gilmore}, Vol.~5, 115

\bibitem[{{Kuznetsov} \& {Kolotkov}(2021)}]{Kuznetsov2021-912}
{Kuznetsov}, A.~A. \& {Kolotkov}, D.~Y. 2021, \apj, 912, 81

\bibitem[{{Lada} \& {Lada}(2003)}]{Lada2003-41}
{Lada}, C.~J. \& {Lada}, E.~A. 2003, \araa, 41, 57

\bibitem[{{Lumsden} {et~al.}(2013){Lumsden}, {Hoare}, {Urquhart}, {Oudmaijer}, {Davies}, {Mottram}, {Cooper}, \& {Moore}}]{Lumsden2013-208}
{Lumsden}, S.~L., {Hoare}, M.~G., {Urquhart}, J.~S., {et~al.} 2013, \apjs, 208, 11

\bibitem[{{Machida} \& {Matsumoto}(2012)}]{Machida2012-421}
{Machida}, M.~N. \& {Matsumoto}, T. 2012, \mnras, 421, 588

\bibitem[{{Ma{\'\i}z Apell{\'a}niz} {et~al.}(2007){Ma{\'\i}z Apell{\'a}niz}, {Walborn}, {Morrell}, {Niemela}, \& {Nelan}}]{Maiz2007-660}
{Ma{\'\i}z Apell{\'a}niz}, J., {Walborn}, N.~R., {Morrell}, N.~I., {Niemela}, V.~S., \& {Nelan}, E.~P. 2007, \apj, 660, 1480

\bibitem[{{Marks} {et~al.}(2012){Marks}, {Kroupa}, {Dabringhausen}, \& {Pawlowski}}]{Marks2012-422}
{Marks}, M., {Kroupa}, P., {Dabringhausen}, J., \& {Pawlowski}, M.~S. 2012, \mnras, 422, 2246

\bibitem[{{Megeath} {et~al.}(2016){Megeath}, {Gutermuth}, {Muzerolle}, {Kryukova}, {Hora}, {Allen}, {Flaherty}, {Hartmann}, {Myers}, {Pipher}, {Stauffer}, {Young}, \& {Fazio}}]{Megeath2016-151}
{Megeath}, S.~T., {Gutermuth}, R., {Muzerolle}, J., {et~al.} 2016, \aj, 151, 5

\bibitem[{{Meynet} \& {Maeder}(2000)}]{Meynet2000-361}
{Meynet}, G. \& {Maeder}, A. 2000, \aap, 361, 101

\bibitem[{{Mezger} \& {Henderson}(1967)}]{Mezger1967-147}
{Mezger}, P.~G. \& {Henderson}, A.~P. 1967, \apj, 147, 471

\bibitem[{{Molinari} {et~al.}(2010){Molinari}, {Swinyard}, {Bally}, {Barlow}, {Bernard}, {Martin}, {Moore}, {Noriega-Crespo}, {Plume}, {Testi}, {Zavagno}, {Abergel}, {Ali}, {Anderson}, {Andr{\'e}}, {Baluteau}, {Battersby}, {Beltr{\'a}n}, {Benedettini}, {Billot}, {Blommaert}, {Bontemps}, {Boulanger}, {Brand}, {Brunt}, {Burton}, {Calzoletti}, {Carey}, {Caselli}, {Cesaroni}, {Cernicharo}, {Chakrabarti}, {Chrysostomou}, {Cohen}, {Compiegne}, {de Bernardis}, {de Gasperis}, {di Giorgio}, {Elia}, {Faustini}, {Flagey}, {Fukui}, {Fuller}, {Ganga}, {Garcia-Lario}, {Glenn}, {Goldsmith}, {Griffin}, {Hoare}, {Huang}, {Ikhenaode}, {Joblin}, {Joncas}, {Juvela}, {Kirk}, {Lagache}, {Li}, {Lim}, {Lord}, {Marengo}, {Marshall}, {Masi}, {Massi}, {Matsuura}, {Minier}, {Miville-Desch{\^e}nes}, {Montier}, {Morgan}, {Motte}, {Mottram}, {M{\"u}ller}, {Natoli}, {Neves}, {Olmi}, {Paladini}, {Paradis}, {Parsons}, {Peretto}, {Pestalozzi}, {Pezzuto}, {Piacentini}, {Piazzo}, {Polychroni}, {Pomar{\`e}s}, {Popescu}, {Reach}, {Ristorcelli},
  {Robitaille}, {Robitaille}, {Rod{\'o}n}, {Roy}, {Royer}, {Russeil}, {Saraceno}, {Sauvage}, {Schilke}, {Schisano}, {Schneider}, {Schuller}, {Schulz}, {Sibthorpe}, {Smith}, {Smith}, {Spinoglio}, {Stamatellos}, {Strafella}, {Stringfellow}, {Sturm}, {Taylor}, {Thompson}, {Traficante}, {Tuffs}, {Umana}, {Valenziano}, {Vavrek}, {Veneziani}, {Viti}, {Waelkens}, {Ward-Thompson}, {White}, {Wilcock}, {Wyrowski}, {Yorke}, \& {Zhang}}]{Molinari2010}
{Molinari}, S., {Swinyard}, B., {Bally}, J., {et~al.} 2010, \aap, 518, L100

\bibitem[{{Motte} {et~al.}(2018){Motte}, {Bontemps}, \& {Louvet}}]{Motte2018-56}
{Motte}, F., {Bontemps}, S., \& {Louvet}, F. 2018, \araa, 56, 41

\bibitem[{{Oey} \& {Clarke}(2005)}]{Oey2005-620}
{Oey}, M.~S. \& {Clarke}, C.~J. 2005, \apjl, 620, L43

\bibitem[{{Purcell} {et~al.}(2013){Purcell}, {Hoare}, {Cotton}, {Lumsden}, {Urquhart}, {Chandler}, {Churchwell}, {Diamond}, {Dougherty}, {Fender}, {Fuller}, {Garrington}, {Gledhill}, {Goldsmith}, {Hindson}, {Jackson}, {Kurtz}, {Mart{\'\i}}, {Moore}, {Mundy}, {Muxlow}, {Oudmaijer}, {Pandian}, {Paredes}, {Shepherd}, {Smethurst}, {Spencer}, {Thompson}, {Umana}, \& {Zijlstra}}]{Purcell2013-205}
{Purcell}, C.~R., {Hoare}, M.~G., {Cotton}, W.~D., {et~al.} 2013, \apjs, 205, 1

\bibitem[{{Rubin}(1968)}]{Rubin1968-154}
{Rubin}, R.~H. 1968, \apj, 154, 391

\bibitem[{{S{\'a}nchez-Monge} {et~al.}(2013){S{\'a}nchez-Monge}, {Beltr{\'a}n}, {Cesaroni}, {Fontani}, {Brand}, {Molinari}, {Testi}, \& {Burton}}]{Sanchez-Monge2013-550}
{S{\'a}nchez-Monge}, {\'A}., {Beltr{\'a}n}, M.~T., {Cesaroni}, R., {et~al.} 2013, \aap, 550, A21

\bibitem[{{Schuller} {et~al.}(2009){Schuller}, {Menten}, {Contreras}, {Wyrowski}, {Schilke}, {Bronfman}, {Henning}, {Walmsley}, {Beuther}, {Bontemps}, {Cesaroni}, {Deharveng}, {Garay}, {Herpin}, {Lefloch}, {Linz}, {Mardones}, {Minier}, {Molinari}, {Motte}, {Nyman}, {Reveret}, {Risacher}, {Russeil}, {Schneider}, {Testi}, {Troost}, {Vasyunina}, {Wienen}, {Zavagno}, {Kovacs}, {Kreysa}, {Siringo}, \& {Wei{\ss}}}]{Schuller2009-504}
{Schuller}, F., {Menten}, K.~M., {Contreras}, Y., {et~al.} 2009, \aap, 504, 415

\bibitem[{{Schulz} {et~al.}(2015){Schulz}, {Pflamm-Altenburg}, \& {Kroupa}}]{Schulz2015-582}
{Schulz}, C., {Pflamm-Altenburg}, J., \& {Kroupa}, P. 2015, \aap, 582, A93

\bibitem[{{Thies} {et~al.}(2015){Thies}, {Pflamm-Altenburg}, {Kroupa}, \& {Marks}}]{Thies2015-800}
{Thies}, I., {Pflamm-Altenburg}, J., {Kroupa}, P., \& {Marks}, M. 2015, \apj, 800, 72

\bibitem[{{Thompson}(1984)}]{Thompson1984-283}
{Thompson}, R.~I. 1984, \apj, 283, 165

\bibitem[{{Urquhart} {et~al.}(2013){Urquhart}, {Thompson}, {Moore}, {Purcell}, {Hoare}, {Schuller}, {Wyrowski}, {Csengeri}, {Menten}, {Lumsden}, {Kurtz}, {Walmsley}, {Bronfman}, {Morgan}, {Eden}, \& {Russeil}}]{Urquhart2013-435}
{Urquhart}, J.~S., {Thompson}, M.~A., {Moore}, T.~J.~T., {et~al.} 2013, \mnras, 435, 400

\bibitem[{{Urquhart} {et~al.}(2022){Urquhart}, {Wells}, {Pillai}, {Leurini}, {Giannetti}, {Moore}, {Thompson}, {Figura}, {Colombo}, {Yang}, {K{\"o}nig}, {Wyrowski}, {Menten}, {Rigby}, {Eden}, \& {Ragan}}]{Urquhart2022-510}
{Urquhart}, J.~S., {Wells}, M.~R.~A., {Pillai}, T., {et~al.} 2022, \mnras, 510, 3389

\bibitem[{{Weidner} \& {Kroupa}(2004)}]{Weidner2004-348}
{Weidner}, C. \& {Kroupa}, P. 2004, \mnras, 348, 187

\bibitem[{{Weidner} \& {Kroupa}(2006)}]{Weidner2006-365}
{Weidner}, C. \& {Kroupa}, P. 2006, \mnras, 365, 1333

\bibitem[{{Weidner} {et~al.}(2013){Weidner}, {Kroupa}, \& {Pflamm-Altenburg}}]{Weidner2013-434}
{Weidner}, C., {Kroupa}, P., \& {Pflamm-Altenburg}, J. 2013, \mnras, 434, 84

\bibitem[{{Wells} {et~al.}(2022){Wells}, {Urquhart}, {Moore}, {Browning}, {Ragan}, {Rigby}, {Eden}, \& {Thompson}}]{Wells2022-516}
{Wells}, M.~R.~A., {Urquhart}, J.~S., {Moore}, T.~J.~T., {et~al.} 2022, \mnras, 516, 4245

\bibitem[{{Yan} {et~al.}(2017){Yan}, {Jerabkova}, \& {Kroupa}}]{Yan2017-607}
{Yan}, Z., {Jerabkova}, T., \& {Kroupa}, P. 2017, \aap, 607, A126

\bibitem[{{Yan} {et~al.}(2023){Yan}, {Jerabkova}, \& {Kroupa}}]{Yan2023-670}
{Yan}, Z., {Jerabkova}, T., \& {Kroupa}, P. 2023, \aap, 670, A151

\bibitem[{{Zhou} {et~al.}(2024{\natexlab{a}}){Zhou}, {Kroupa}, \& {Dib}}]{Zhou2024arXiv240809867Z}
{Zhou}, J.~W., {Kroupa}, P., \& {Dib}, S. 2024{\natexlab{a}}, arXiv e-prints, arXiv:2408.09867

\bibitem[{{Zhou} {et~al.}(2024{\natexlab{b}}){Zhou}, {Kroupa}, \& {Dib}}]{Zhou2024-688}
{Zhou}, J.-w., {Kroupa}, P., \& {Dib}, S. 2024{\natexlab{b}}, \aap, 688, L19

\end{thebibliography}


\appendix

\section{The theoretical $m_{\rm max}-M_{\rm ecl}$ relation}\label{mM}
\begin{figure}
\centering
\includegraphics[width=0.45\textwidth]{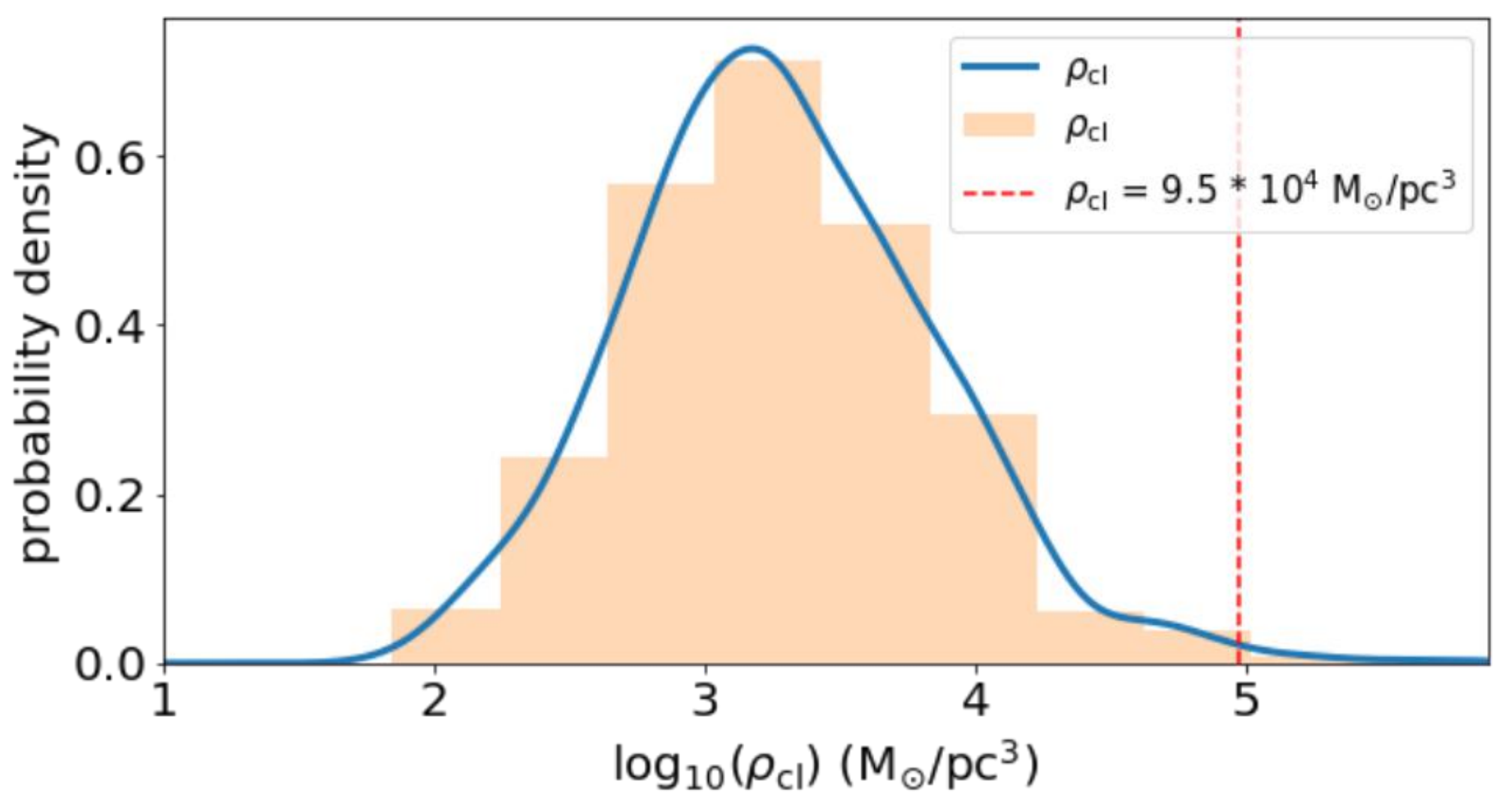}
\caption{The volume density distribution of ATLASGAL clumps with HII regions.}
\label{den}
\end{figure}

The stellar IMF, $\xi_{\mathrm{\star}}(m)$, is canonical \citep{Kroupa2001-322}:
\begin{equation}\label{eq:xi_star}
\xi_{\mathrm{\star}}(m,M) =
\begin{cases} 
0, & m<0.08~\mathrm{M}_{\odot}, \\
2k_{\mathrm{\star}} m^{-1.3}, & 0.08~\mathrm{M}_{\odot} \leqslant m<0.5~\mathrm{M}_{\odot}, \\ 
k_{\mathrm{\star}} m^{-2.3}, & 0.5~\mathrm{M}_{\odot} \leqslant m<1\mathrm{M}_{\odot}, \\
k_{\mathrm{\star}} m^{-\alpha_3}, & 1~\mathrm{M}_{\odot} \leqslant m<m_{\mathrm{max}}(M), \\
0, & m_{\mathrm{max}}(M) \leqslant m,
\end{cases}
\end{equation}
where $\alpha_3=2.3$ is the constant Salpeter-Massey index for the invariant canonical IMF but will change for larger $\rho_{\mathrm{cl}}$ (the density of star-forming clump) to account for IMF variation under star-burst conditions \citep{Elmegreen2003-338, Dib2007-661,Dabringhausen2009-394,Marks2012-422}.
0.08 M$_{\odot}$ in Eq.~\ref{eq:xi_star} is about the lower mass limit of stars \citep{Thies2015-800}. 
The mass conservation of the embedded cluster gives
\begin{equation}\label{eq:MeclintMstar}
M_{\mathrm{ecl}}=\int_{0.08~\mathrm{M}_{\odot}}^{m_{\mathrm{max}}}m~\xi_{\mathrm{\star}}(m)\,\mathrm{d}m.
\end{equation}
The optimal sampling normalization condition is
\begin{equation}\label{eq:1intMstar}
1=\int_{m_{\mathrm{max}}}^{150~\mathrm{M}_{\odot}}\xi_{\mathrm{\star}}(m)\,\mathrm{d}m,
\end{equation}
where 150 M$_{\odot}$ is the adopted stellar upper mass limit \citep{Weidner2004-348, Figer2005-434, Oey2005-620, Koen2006-365, Maiz2007-660}. When $m_{\mathrm{max}} > 1 \mathrm{M}_{\odot}$,
the optimal sampling normalization condition becomes
\begin{equation}\label{eq:1intMstar}
1=\int_{m_{\mathrm{max}}}^{150~\mathrm{M}_{\odot}} k_{\mathrm{\star}} m^{-\alpha_3} \,\mathrm{d}m.
\end{equation}

For larger $\rho_{\mathrm{cl}}$, $\xi_{\mathrm{\star}}(m)$ becomes top-heavy where a $\alpha_3(\rho_{\mathrm{cl}})$ relation is adopted from \citet{Marks2012-422}:
\begin{equation}\label{eq:alpha_3}
\alpha_3=
\begin{cases} 
2.3, & \rho_{\mathrm{cl}}<9.5\times 10^4, \\
1.86-0.43\log_{10}(\rho_{\mathrm{cl}}/10^6), & \rho_{\mathrm{cl}} \geq 9.5\times 10^4.
\end{cases}
\end{equation}
Here 
\begin{equation}\label{eq:rho_clD}
\rho_{\mathrm{cl}}=3M_{\mathrm{cl}}/4\pi r_{\rm h}^3
\end{equation}
in the unit of $[\mathrm{M}_{\odot}/\mathrm{pc}^3]$ is the clump density when the embedded cluster is forming.
For $M_{\mathrm{cl}}$ and $r_{\rm h}$ in the equation, we take the clump's mass and radius calculated in \citet{Urquhart2022-510}.
As shown in Fig.\ref{den}, for ATLASGAL clumps with HII regions (HII-clumps), generally, they have $\rho_{\mathrm{cl}}<9.5\times 10^4$. Thus, we take $\alpha_3$=2.3 in this work. 

Setting $x \equiv m_{\mathrm{max}}$ and solving the above equations, we obtain
\begin{equation}
\begin{split}
& M_{\mathrm{ecl}}=\int_{0.08~\mathrm{M}_{\odot}}^{x}m~\xi_{\mathrm{\star}}(m)\,\mathrm{d}m\\ &
=\int_{0.08~\mathrm{M}_{\odot}}^{0.5~\mathrm{M}_{\odot}}m~2k_{\mathrm{\star}} m^{-1.3}\,\mathrm{d}m+\int_{0.5~\mathrm{M}_{\odot}}^{x}m~k_{\mathrm{\star}} m^{-2.3}\,\mathrm{d}m
\\ & \approx 
\frac{5.37}{\frac{0.77}{x^{1.3}} - 0.001} - \frac{3.33}{x^{0.3} \left(\frac{0.77}{x^{1.3}} - 0.001\right)}
\end{split}
\end{equation}

\section{Robust of the model}
\begin{figure}
\centering
\includegraphics[width=0.475\textwidth]{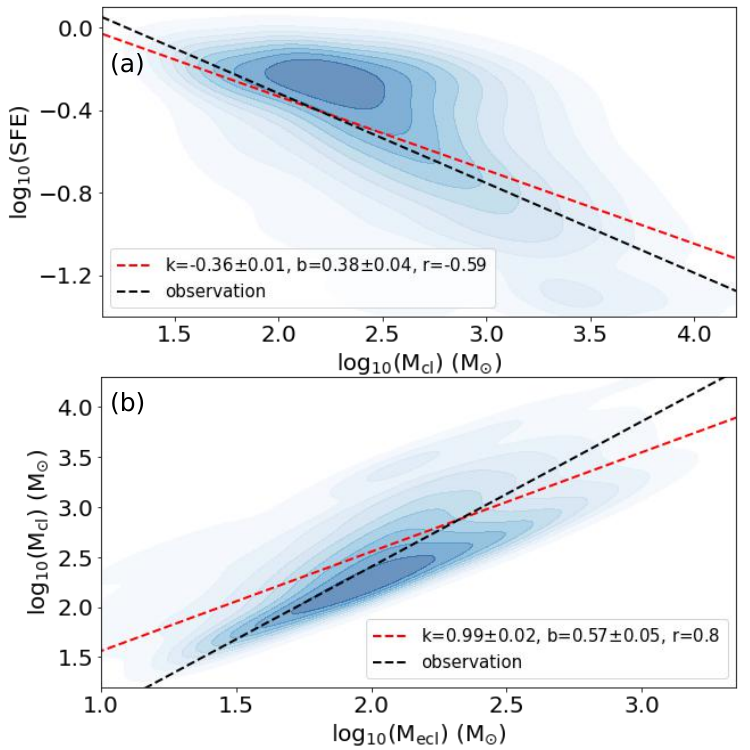}
\caption{Same as Fig.\ref{sfe-m}. But the star formation efficiency is derived from the case of the second optimal combination, as described in Sec.\ref{s-sfe}.}
\label{sfe-ms}
\end{figure}
\begin{figure}
\centering
\includegraphics[width=0.475\textwidth]{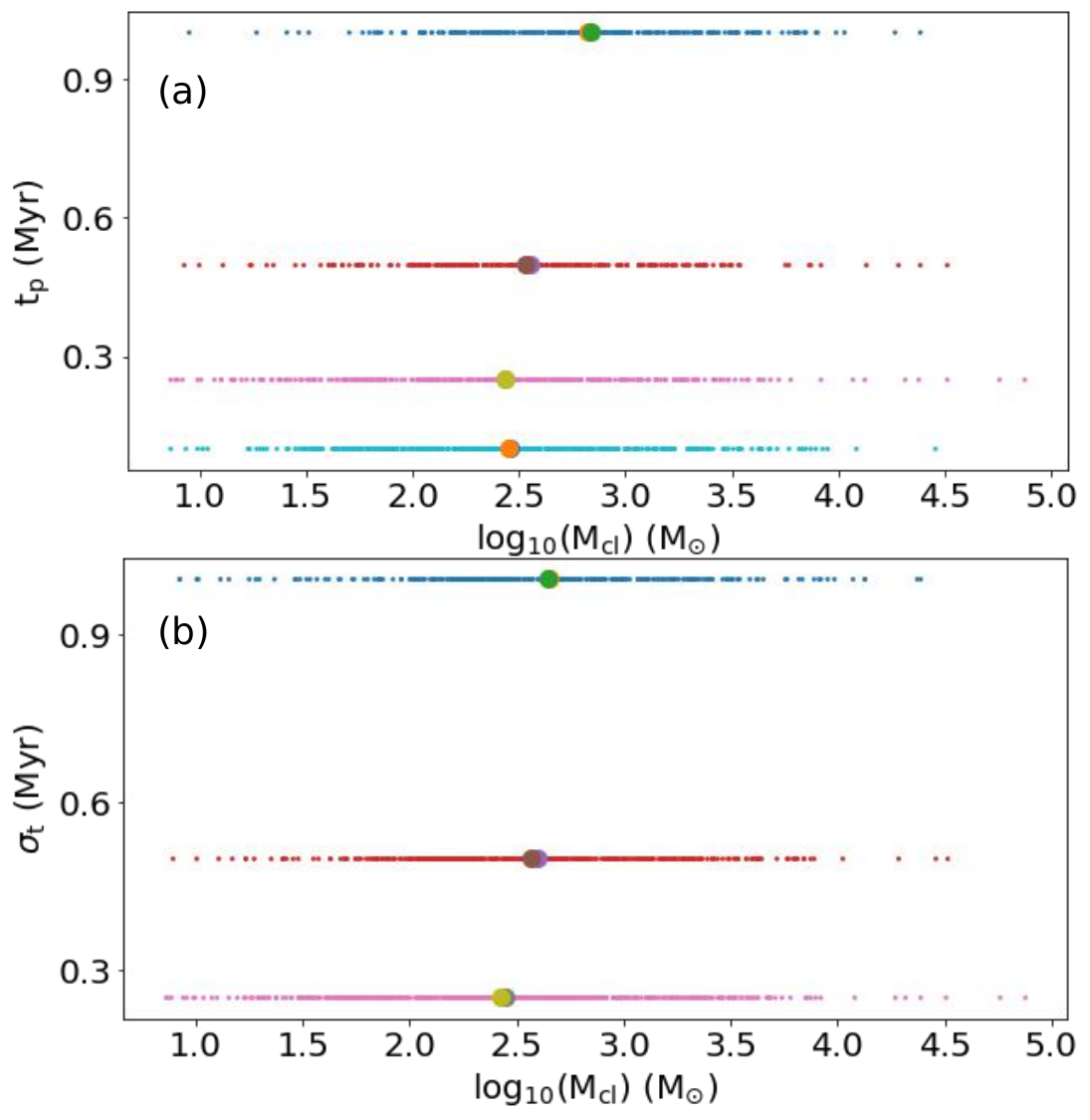}
\caption{Same as Fig.\ref{sfh}, but for the case of the second optimal combination, as described in Sec.\ref{s-sfe}.}
\label{sfh1}
\end{figure}

\end{document}